\documentclass[sigconf,nonacm]{acmart}
\usepackage{array}       
\usepackage{booktabs}    
\usepackage{multirow}    
\usepackage{graphicx}    
\usepackage{amsmath}     
\usepackage{float} 
\usepackage{textcomp}

\usepackage{framed}
\usepackage{alltt}

\usepackage{CJK}   

\usepackage{flafter} 
\usepackage{stfloats}

\newcommand{\rev}[1]{#1}
\newcommand{\add}[1]{#1}
\newcommand{\minor}[1]{#1}

\begin{document}
\begin{CJK}{UTF8}{gbsn}

\title[Characterizing Scam-Driven Human Trafficking and Online Community Responses on RedNote]{Characterizing Scam-Driven Human Trafficking Across Chinese Borders and Online Community Responses on RedNote}

\author{Jiamin Zheng}
\affiliation{%
  \institution{University of Edinburgh}
  \city{Edinburgh}
  \country{United Kingdom}}
\email{jiamin.zheng@ed.ac.uk}

\author{Yue Deng}
\affiliation{%
  \institution{Department of Computer Science and Engineering, The Hong Kong University of Science and Technology}
  \city{Hong Kong}
  \country{China}}
\email{ydengbi@connect.ust.hk}
\authornote{This work was conducted while the author was a visiting PhD scholar at the Max Planck Institute for Security and Privacy.}

\author{Jessica Chen}
\affiliation{%
  \institution{University of Edinburgh}
  \city{Edinburgh}
  \country{United Kingdom}}
\email{J.Chen-265@sms.ed.ac.uk}

\author{Shujun Li}
\affiliation{%
  \institution{University of Kent}
  \city{Canterbury}
  \country{United Kingdom}}
\email{s.j.li@kent.ac.uk}

\author{Yixin Zou}
\affiliation{%
  \institution{Max Planck Institute for Security and Privacy}
  \city{Bochum}
  \country{Germany}}
\email{yixin.zou@mpi-sp.org}

\author{Jingjie Li}
\affiliation{%
  \institution{University of Edinburgh}
  \city{Edinburgh}
  \country{United Kingdom}}
\email{jingjie.li@ed.ac.uk}


\begin{abstract}
A new form of human trafficking has emerged across Chinese borders, where individuals are lured to Southeast Asia with fraudulent job offers and then coerced into operating online scams. Despite its massive economic and human toll, this scam-driven trafficking remains underexplored in academic research. Through qualitative analysis of 158 RedNote posts, we examined how Chinese online communities respond to this threat. Our findings reveal that perpetrators exploit cultural ties to recruit victims for cybercriminal roles within self-sustaining compounds, using sophisticated manipulation tactics. Survivors face serious reintegration barriers, including family rejection, as the cultural values that enable trafficking also hinder their recovery. While communities present protective strategies, efforts are complicated by doubts about the reliability of support and cross-border coordination. We discuss key implications for prevention, platform governance, and international cooperation against scam-driven trafficking.
Warning: This paper contains descriptions of physical, psychological, and sexual abuse.
\end{abstract}


\maketitle
\noindent\textit{Accepted at CHI 2026. DOI: 10.1145/3772318.3791786}
\vspace{0.5em}

\section{Introduction}

The rapid expansion of digital technologies has transformed human trafficking into hybrid forms of exploitation that extend beyond traditional labor and sexual coercion. Increasingly, traffickers deploy technology-enabled tactics that are broader, faster, and harder to detect. This phenomenon, often described as ``cyber slavery''~\cite{liu2023cyber}, involves individuals deceived by promises of lucrative jobs, trafficked across borders, and forced to work in scam compounds running large-scale online fraud operations. 

Recent reports highlighted China as one of the primary sources of human trafficking. In 2023 alone, Chinese authorities announced the arrest of over 50,000 fraud suspects and the disruption of organized networks in Myanmar's border region~\cite{Lin2025_ChineseCrimeSEA}. A report published by the Office of the United Nations High Commissioner for Human Rights (OHCHR) in 2023 estimated that hundreds of thousands of people are trapped in cyber-scam compounds across Southeast Asian countries near Chinese borders, such as Cambodia, Myanmar, and Laos, generating approximately \$43.8 billion annually for criminal networks~\cite{ohchr2023-online-scam-sea}. At least 120,000 individuals were reportedly forced into scam labor in Myanmar and another 100,000 in Cambodia~\cite{ohchr2023-online-scam-sea}. This modern form of human trafficking blurs the victim-perpetrator boundary~\cite{Wang2024}, as trafficked individuals are compelled to commit crimes against others while being victims themselves.

Technology plays a dual role in this ecosystem. Traffickers exploit mobile phones, social media platforms, and encrypted communication tools to recruit, control, and monitor victims~\cite{preiss2022digital, stephenson2025digital}. However, with increasing usage of social media~\cite{saud2020usage}, it provides vital spaces where survivors, families, and concerned users share warnings, coordinate rescue efforts, and build collective knowledge about trafficking operations, mirroring the broader online safety and well-being research showing how users and communities develop narratives around experiences of risks and protective strategies and provide peer support~\cite{kou17Conspiracy, zhang2018online}. 

Despite its severity, scam-driven human trafficking remains critically underexplored in the academic literature. \rev{Existing work in this setting has largely examined institutional responses and broader sex and labor exploitation~\cite{stockl2021human, oram2016prevalence}, along with compound structures, governance dynamics, and legal responses~\cite{jesperson2023trafficking, ohchr2023-online-scam-sea, usip2024mekong}. Empirical studies show that traffickers recruit educated, digitally skilled victims through fake job advertisements, operate within layered criminal networks enabled by weak governance, and force victims to conduct scripted online fraud under violent enforcement regimes~\cite{luong2024understanding, luong2025simple, jesperson2023trafficking, lazarus2025assessing}.} \rev{R}egional studies have traced how cyberfraud industries have migrated from China into Southeast Asia, supported by weaker governance, complicit local authorities, and money-laundering infrastructures~\cite{franceschini2023compound, Loughlin2024, Chen2024}. 

\rev{Building on prior work examining institutional responses or isolated survivor accounts, we focus on} how victims and communities themselves perceive risks and develop protective strategies in the digital space of scam-driven human trafficking.
\rev{From this lens, we can understand how different stakeholders exchange information and support across the full trafficking lifecycle, as well as the cultural and socio-technical dynamics that shape community responses when institutional interventions prove inadequate. Specifically, we address four research questions:}

\begin{itemize}
    \item \textbf{RQ1:} What \textit{recruitment tactics} for scam-driven human trafficking do RedNote users recognize?
    \item \textbf{RQ2:} What \textit{exploitation and control mechanisms} in scam-driven human trafficking are shared by RedNote users?
    \item \textbf{RQ3:} What \textit{post-trafficking outcomes and reintegration challenges} are identified by RedNote users affected by scam-driven human trafficking? 
    \item \textbf{RQ4:} How do RedNote users share and evaluate \textit{protective strategies} against scam-driven human trafficking?
\end{itemize}

To answer these questions, we conducted a qualitative content analysis based on 158 posts sampled from RedNote, \rev{a popular Chinese social media platform that has recently become a valuable data source for researchers}~\cite{wan-chi25, deng-deng25, yuan2025love}. Our findings \rev{reveal how perpetrators weaponize Chinese kinship ties and filial duties throughout the trafficking lifecycle. During recruitment (\textbf{RQ1}), scammers' strategies extend beyond generic social ties identified by prior work~\cite{luong2025simple, lazarus2025assessing}: they exploit trust embedded in kinship and cultural expectations, while the boundary between voluntary participation and coercion becomes blurry for victims. While certain groups are especially vulnerable, including ``left-behind youth''---rural children whose migrant-worker parents leave them with a weak social safety net, even highly skilled individuals are not immune and are specifically targeted in recruitment for multilingual abilities or other competencies valuable to scam operations. Once trafficked (\textbf{RQ2}), survivors endure continuous, multi-layered abuse in scam compounds, while operators further weaponize kinship and cultural ties to extract ransoms from families. Upon escape, survivors seeking reintegration (\textbf{RQ3}) face family rejection, uncertain legal consequences, and public shaming for being ``greedy'' or ``gullible,'' in stark contrast to the sympathy typically afforded to abuse survivors~\cite{pevac2022online}. When institutional channels prove inadequate, RedNote users develop their own protective strategies (\textbf{RQ4}), including grassroots rescue networks and targeted warnings for vulnerable groups. These community efforts raise broader awareness and address gaps left by formal support systems.} 

Situating our work within the HCI literature~\cite{wash2020experts, razaq2021we, randazzo2023if, starbird2011voluntweeters}, we discuss pathways to protect survivors of scam-driven human trafficking, including culturally and socially informed digital safenets, content moderation on traumatizing topics and narratives, and efforts related to cross-border rescue.

For the rest of the paper, we review related work on human trafficking and online safety discourse (\S\ref{sec:related}), then describe our qualitative methodology for analyzing RedNote posts (\S\ref{sec:method}). We present findings organized by our four research questions (\S\ref{sec:findings}) and discuss implications for culturally informed interventions, platform design, and cross-border support systems (\S\ref{sec:discussion}).

\section{Background and Related Work}
\label{sec:related}
Below, we introduce the background for our study on scam-driven human trafficking, a pressing threat that involves deceiving victims through false employment promises and coercing them into cybercrime operations across international borders. 

\subsection{Exploitation in Human Trafficking}

Exploitation in human trafficking has historically been framed around sexual and labor exploitation, including physical violence, psychological manipulation, and economic coercion~\cite{kim2010coercion, ollus2015regulating}. More than half of survivors report physical or sexual abuse, while psychological coercion creates emotional dependencies that isolate victims from support networks~\cite{stockl2021human, oram2016prevalence, merodio2020they}. Economic mechanisms such as debt bondage, confiscation of documents, and wage withholding further entrench dependency and limit mobility~\cite{stephenson2025digital}. These practices are shaped by inequalities, with marginalized groups disproportionately targeted and subjected to slavery-like conditions~\cite{limoncelli2009trafficking, shepherd2022organizing}. Beyond sex and labor exploitation, trafficking also involves forms of exploitation such as domestic servitude, organ trafficking, and criminal exploitation, though these remain comparatively under-researched~\cite{cockbain2019human}. Recent work has highlighted that victims are increasingly coerced into illicit activities such as cybercrime, which blurs the boundary between victims and perpetrators~\cite{Wang2024}. 

With the rapid development of China's economy and societal transformation, fraud has also exhibited a trend towards digitalization. As Chinese authorities intensify cross-border crackdowns and repatriations with Southeast Asian partners, Chinese nationals remain a major share of both victims and perpetrators in scam compounds~\cite{StateCouncil2021ActionPlan}. Victims deceived with fraudulent job offers are coerced into cyber-fraud operations within heavily guarded compounds, facing strict quotas, extended hours, and violent punishments for non-compliance~\cite{hrc2023guidance, hrc2022cyberslaver}. Testimonies have shown that traffickers train victims to fabricate online identities and carry out ``pig-butchering'' scams that combine romantic persuasion with fraudulent investment schemes~\cite{sarkar2024bidirectional, Wang2024, whitty-cbs12, oak-soups25}. \citeauthor{Wang2024}~\cite{Wang2024} showed how trafficked workers undergo coerced identity shifts that sustain their dual status as both exploited and criminalized. 

While recent research has relied primarily on interviews, NGO reports, and legal proceedings, less attention has been paid to how Chinese survivors and communities interpret these exploitative practices in digital spaces. Our study addresses this gap by examining how users recognize, describe, and evaluate the mechanisms of scam-driven human trafficking through RedNote data.

\subsection{Human Trafficking in the Digital Era}

Human trafficking is defined as the exploitation of individuals through force, fraud, or coercion for involuntary labor or sexual activity~\cite{UNODC2024trafficking}. \citeauthor{kleemans2014human}~\cite{kleemans2014human} conceptualized trafficking as a linear process involving recruitment, transportation, and exploitation. Digital technologies have transformed the methods traffickers use to recruit, control, and exploit victims. Mobile phones, social media platforms, and chat applications are central tools for contacting victims, advertising services, and coordinating transactions in sex and labor trafficking contexts~\cite{latonero2012technology}. Technology also amplifies coercion beyond traditional forms of physical and psychological abuse. As \citeauthor{stephenson2025digital}~\cite{stephenson2025digital} demonstrated, human traffickers exert real-time control of survivors via location trackers such as Apple's FindMy, covert monitoring apps, and contact surveillance. 

While the existing scholarship on sex and labor trafficking emphasizes recruitment through social media and escort websites, scam-driven trafficking in Asia operates through ``forced criminality,'' wherein victims are deceived into cyber-fraud operations where they are compelled to use digital tools to target new victims, effectively turning coercion into a scalable digital labor system, a phenomenon journalists referred to as ``cyber slavery''~\cite{hingston2023forced, hrc2022cyberslaver, sarkar2024bidirectional}. \rev{These operations are concentrated in border regions such as Myawaddy in Myanmar, where numerous scam compounds have been documented.} \add{Regional work traces how cyber-fraud industries moved from China to Southeast Asia, exploiting weak governance and legal loopholes. \citeauthor{franceschini2023compound}~\cite{franceschini2023compound} detailed this shift through a case study of Sihanoukville, Cambodia, while others described protection by political elites in Cambodia and military-linked actors in Myanmar~\cite{Loughlin2024, usip2024mekong}. Institutional reports frame scam compounds as part of transnational organized crime. In 2023, OHCHR~\cite{ohchr2023-online-scam-sea} called for cross-border cooperation and emphasized the non-punishment principle for coerced victims.}

\add{While news reports highlight rapid growth in scam compounds and mass trafficking into cyber-fraud hubs with statistics~\cite{McPherson2025, Ratcliffe2025GuardianMyanmarScam, Xinhua2025Feb, Xinhua2023Aug}, empirical studies add more granular views. \citeauthor{luong2025simple}~\cite{luong2025simple} analyzed Vietnamese case files and police interviews and showed that traffickers target educated victims through fake job adverts in layered criminal networks. \citeauthor{jesperson2023trafficking}~\cite{jesperson2023trafficking} studied eight Southeast Asian countries and identified enabling institutional and economic conditions, including targeting highly educated workers, reversed migration flows from wealthier to poorer countries, and dual victimization of both trafficked workers and scam targets. \citeauthor{lazarus2025assessing}~\cite{lazarus2025assessing} drew from a Bangladeshi survivor's testimony to outline six stages from recruitment to escape and show how victims are coerced into scripted online fraud.}

\rev{In terms of institutional support for human trafficking survivors, counter-trafficking efforts in the US rely on hotlines, communication apps, and online awareness campaigns that allow victims and families to seek help remotely and enable NGO coordination~\cite{feeney2024evaluation}, while institutional support for Chinese human-trafficking victims heavily relies on law enforcement and official anti-scam apps~\cite{han_new_2019, xue2023internet}. However, rising evidence has shown that people have found these channels limited or inaccessible, thus turning to social media to share information, mobilize support, and seek help directly under public safety crisis~\cite{peterson2019officialsystems, saud2020usage, bbcNews-25}.} This trend highlights that social media is not only a site of risk, where traffickers recruit and exploit, but also a critical source of support and community resilience.

\add{For methodologies, news coverage tends to focus on high-profile cases and rescue operations, which could be subject to media ideology and political environments~\cite{stroud2008media}. Past academic research has documented trafficking organizations, economic impacts, and survivor experiences, mostly through interviews and official reports outside China~\cite{jesperson2023trafficking, luong2024understanding, luong2025simple, lazarus2025assessing}. Chinese Internet users' own perspectives on how they understand risks, exchange support, and navigate reintegration in everyday digital spaces remain under-studied, despite being a major victim group and a primary trafficker nationality~\cite{luong2025simple}. Our study addresses this gap by analyzing posts on RedNote, a Chinese social media platform where users actively discuss safety risks, help-seeking, and reintegration under crisis in their own narratives.}

\subsection{\rev{User Perception and Online Safety Discourse}}


Online communities develop their understandings of risk through shared narratives, warnings, and interpretive frameworks that often operate alongside or in tension with official safety guidelines~\cite{kou17Conspiracy, nagar2012collective, wei2024understanding}. Research on adolescent online safety has demonstrated how users interpret linguistic and emotional cues to distinguish safe from unsafe interactions~\cite{ali2022understanding}, while research on family dynamics has highlighted tensions between trust, autonomy, and protective strategies~\cite{hartikainen2016discourses}. These perspectives underscore that users are not passive targets but actively co-construct safety practices in digital spaces. Research in security and privacy has shown how online discussions can reveal everyday safety concerns and coping strategies, from smart homes to software development~\cite{vafa-sp25, ali-sp25, li-sp23, bellini2021so, li2021developers, tahaei2020understanding, whiting2019online}. \add{In particular, human-centered security research has examined how users detect online scams leveraging URLs and browser security indicators~\cite{dhamija2006phishing, sheng2010phish, egelman2008youve, sunshine2009crying} and the barriers users encounter during their sense-making process~\cite{downs2007behavioral, wash2020experts, nthala2021nonexperts}. Recent work has also leveraged online forums to analyze scam-related discourse, showing that communities collaboratively dissect fraudulent tactics, reconstruct scam lifecycles, and share step-by-step prevention guidance~\cite{oak2025discourse}. Community resources can update faster than institutional advisories and offer granular, platform-specific advice.} These investigations highlight how personal narratives on digital platforms often serve as user-driven strategies for interpreting, flagging, or resisting perceived harms.

\rev{Cultural and political contexts further influence online safety discourses,  particularly for sensitive or controversial topics~\cite{zhou-chi17, tam-chi23, deng-deng25}. In the Chinese context, \citeauthor{deng-deng25} showed how filial piety motivated people to seek online safety advice for parents~\cite{deng-deng25}, and \citeauthor{he2025living} discovered that social media commodified the safety concerns of Chinese women~\cite{he2025living}. This is also manifested in the moderation policies of online platforms, which drove Chinese users to use indirect expression and re-appropriate hashtags when articulating threats and maintaining online safe spaces~\cite{pater2016harassment, wu2021hidden, wan-chi25, mayworm2024content}; while weaker moderation can also push communities to create local safety norms and workaround practices~\cite{sabri2023challenges}. These patterns are relevant to trafficking-related discussions, which share features with other forms of online exploitation, such as grooming and sextortion~\cite{razi2023sliding}.}

\rev{Prior work explains how individuals actively interpret risks, co-construct safety narratives and adapt to platform governance. However, much existing research in HCI focuses on phishing and general forms of online fraud in Western contexts. Emerging studies in China have begun to analyze social media users' discourse and responses to both public and individual safety crises, including the COVID-19 pandemic~\cite{lu-cscw21}, mental health issues~\cite{zhang2018online}, and gender-specific harms~\cite{he2025living}, highlighting how these dynamics are shaped by unique cultural, social, and technological environments in China. Scam-driven human trafficking in China, however, extends far beyond individual safety risks in contexts like email-based phishing, and instead represents a systematic public crisis involving sophisticated digital and physical fraud schemes, the exploitation of digital labor, and ambiguous online governance spaces and support systems. Despite its complexity and societal impact, there remains a lack of holistic research examining how Chinese Internet users respond to different phases of scam-driven human trafficking and collectively construct protective discourses in this rapidly evolving, sensitive, and often contentious context. Addressing this gap motivates our study.}

\section{Method}
\label{sec:method}
To understand human trafficking driven by scams from a user-centric perspective, we collected relevant posts on RedNote, a major Chinese social network, and conducted qualitative content analysis to uncover insights at different stages of scam-driven human trafficking.

\subsection{Data Source}

We chose social media platforms as our data source for studying scam-driven trafficking discourse because they capture diverse user narratives about the full life-cycle of trafficking experiences in an ecologically valid setting. Specifically, we selected RedNote, a rapidly growing Chinese social media platform launched in 2013 that combines features from Instagram and TikTok, supporting lifestyle-oriented content through text, photos, and videos~\cite{sun2023influence, huang2024domesticating}. By 2025, RedNote had over 300 million monthly active users, with a user base largely composed of young people\rev{, who are disproportionately targeted by trafficking~\cite{scmpAlibabaTeams, walby2016study}. RedNote has gained increasing attention in HCI research as the study site for Chinese users on topics related to online safety and discourse analysis of personal experiences~\cite{wan-chi25, deng-deng25,li2025challenges, he2025living}. It particularly aligns with our research objectives as RedNote emphasizes individual perspectives and global experiences of Chinese users within a unified app~\cite{ying_acceptance_2023}, which allows analysis of firsthand accounts of experiences and protection practices regarding cross-border scam-driven human trafficking. In contrast, other platforms such as Douyin (the domestic version of TikTok) maintain separate domestic and international versions or host limited relevant content. Our preliminary exploration indicates that RedNote posts contain richer personal narratives about scam-driven human trafficking. Searches on Weibo, another candidate mainstream social media platform in China returned fewer such narratives and were dominated by official announcements, likely a result of the platform's content moderation policies~\cite{zhu2012tracking, Zhu-sec13}, making it less suitable for our purposes.}

\subsection{Data Collection}

To identify posts relevant to scam-driven human trafficking from RedNote, we collected publicly available content from RedNote using MediaCrawler~\cite{nanmicoder2025mediacrawler}, which queries RedNote's web API to automate keyword-based search and post download. To mitigate biases introduced by personalized recommendations, we used a fresh RedNote account for data collection. RedNote's search feature relies on contextual similarity rather than strict keyword matching~\cite{deng-deng25}.

Our search term selection drew upon the trafficking-related literature, media reports, and online discussions to ensure we covered a diverse topic representation for different aspects and stages of scam-driven human trafficking. We include 26 keywords under four main groups of key terms: (1) core trafficking terminology (``\texttt{human trafficking}''), (2) scam-related terms (``\texttt{high-paying labor}'' and ``\texttt{overseas job scam}''), (3) geographic targets that incorporate specific high-risk regions frequently mentioned in trafficking discourse (``\texttt{Northern Myanmar}'' and ``\texttt{Myawaddy campus}''), and (4) victim experience terminology that captures first-hand narratives (``\texttt{trafficking experience}'' and ``\texttt{abduction experience}''). We started with terms that are broad in scope, formed and tested more specific terms by snowball sampling from the key terms and tags in the collected samples initially. Data collection was completed in January 2025, and a total of 6,639 posts were retrieved.  

We used a comprehensive keyword set to cover more posts from different perspectives, as each search returned around 200 to 300 posts. We collected posts by searching each of the 26 terms. Our data exploration showed that posts containing particular tags about entertainment and lifestyles (e.g., dramas and food recommendations) are largely irrelevant to user narratives of scam-driven human trafficking. Therefore, we excluded posts associated with these tags. As a result, our collected posts include keywords and tags subject to the logic \texttt{(human trafficking OR high-paying labor OR ... OR Cambodian scam)} \texttt{AND NOT} \texttt{(Korean drama OR Daily drama-watching OR ... OR Food and fun)}. We show the list of keywords and excluded tags in Appendices~\ref{appendix:keywords} and \ref{appendix:excluded_tags}, respectively. After removing duplicated entries using each post's unique ID, we retained a dataset that includes 4,499 posts, with post dates ranging from December 2018 to January 2025.

\subsection{Data Filtering and Sampling}

Despite initial scoping using key terms and tags, the dataset still contained posts that were irrelevant to scam-driven human trafficking, e.g., human trafficking without connections to scams. We discussed and established the relevance criteria that a post should contain relevant information about scams and human trafficking, or cross-border human trafficking between China and Southeast Asia that is connected to scam operations. These posts include types such as users' first-hand experiences, commentary on related news, and information-seeking posts. We employed a Chinese LLM (large language model) \texttt{qwen2.5-3b} to complement our manual efforts to identify relevant data points, following prior work that adopts similar semi-automated approaches~\cite{wei2024understanding}.

\minor{We developed a prompt (Appendix~\ref{sec:prompt-llm}) for \texttt{qwen2.5-3b} to identify candidate posts for subsequent manual validation. Final inclusion decisions were made by the authors through close verification. We ran \texttt{qwen2.5-3b} locally over the full corpus of 4,499 posts and retrieved 1,955 candidate posts for manual review.}

\minor{To assess screening quality, we randomly sampled 100 posts from the full corpus (44 predicted as relevant by the model and 56 predicted as irrelevant) and had the first two authors independently annotate relevance as ground truth, using a deliberately permissive inclusion criterion that considers borderline posts as relevant candidates. For example, we treated posts that mention scam-driven human trafficking only via hashtags as relevant candidates for manual filtering. This permissive criterion was designed to evaluate the model's ability to flag any post that plausibly fell within our final inclusion scope. Inter-rater agreement was high (Cohen's $\kappa = 0.98$), indicating consistent application of the criteria. The model achieved 84\% accuracy and a false negative rate of 8.8\% in identifying candidate posts for human validation. Upon closer manual examination, we found that posts missed by the filter model were typically borderline or low-information cases (e.g., \texttt{\#human\_trafficker} appeared in the hashtags, but the body text discussed unrelated topics or lacked substantive detail). These missed posts therefore would not contribute confident, substantive information even if included in the qualitative analysis.}


Using \texttt{qwen2.5-3b}, we obtained a subset of 1,955 candidate posts from our dataset. \add{We drew random samples from the candidate posts and then jointly validated and further refined the samples by the first two authors as the coders. We removed posts unrelated to scam-driven human trafficking and excluded those without substantive information (e.g., scam reports with no link to overseas trafficking and scam compounds and domestic missing-person cases without suspected trafficking to scam compounds) and resolved ambiguous cases during team meetings.} As a result, our final data sampled and coded for analysis included 158 posts, during which we reached data saturation~\cite{saunders2018saturation} when establishing our codebook used in thematic analysis using the first 93 posts sampled.

\subsection{Qualitative Content Analysis}

We conducted open coding for our thematic analysis. Our initial codes are built on \citeauthor{kleemans2014human}'s linear model of recruitment, transportation,  and exploitation in human trafficking~\cite{kleemans2014human}. Nevertheless, existing models missed the details in the user narratives of scam-driven trafficking documented on social media. \add{To derive fine-grained codes and insights, the first author first familiarized themselves with the dataset and developed initial codes and themes. The second author jointly coded the dataset and contributed to the code refinement and interpretations. The two coders worked closely together, meeting weekly to reconcile disagreements and ambiguities in the codes and definitions. The wider team was involved throughout the process to review the codebook and help refine the coding scheme. The first author then reapplied the updates and revisions to the coded data, ensuring consistency.} The two coders established our initial codebook using 60 posts, for which we applied axial coding~\cite{corbin2008basics} to identify associated codes and form a hierarchy. After that, we further coded data and monitored data saturation~\cite{saunders2018saturation}. We observed data saturation using 93 posts, after which we continued to code 65 posts. 

\add{Throughout the analysis, we adopted a reflexive thematic analysis approach~\cite{braun2019reflecting}, treating researchers as active interpreters rather than interchangeable coders applying fixed categories. The varied expertise within our team in HCI, online security and safety, and cybercrime supported a richer understanding of nuances in the data. Following qualitative practice in HCI~\cite{mcdonald2019reliability}, we prioritized iterative discussion and collective sense-making rather than independent coding agreement. Inter-rater reliability metrics are not appropriate for this analytic approach~\cite{braun2019reflecting} because IRR assumes stable categories and would constrain the interpretive flexibility needed to capture the complexity of user narratives. We provide our codebook in Appendix~\ref{sec:codebook}, which includes 52 codes and 91 fine-grained sub-codes.} 



\subsection{Ethical Considerations}
\label{sec:ethics}
Our research was approved by the Ethics Committee of the School of Informatics, University of Edinburgh. We paid careful attention to ensuring ethical research practices, particularly in handling sensitive content and user information, although the posts are already publicly available. We removed all usernames and identifiers when presenting our analysis. We further paraphrased and manually translated our quotes from Chinese to English to minimize re-identification of post authors via direct searches on RedNote. 
\add{Each quoted post in our reported findings is indexed by a unique label (e.g., \texttt{P-e8Y}), where \texttt{P} denotes \emph{Post} and the alphanumeric suffix was randomly generated. These labels serve solely to distinguish between different posts.}
Our analysis focuses on aggregate insights, and we did not attempt to or intend to trace, contact, or profile individual users. To reduce the risk of undue platform strain, we did not collect more than 300 posts per day. The Chinese LLM we used is hosted locally without sharing data with a third party. Our study team members were informed of the topic and concerns for analyzing the potentially triggering content, and we carefully monitored the mental well-being of team members throughout the project. \add{Team members took structured breaks to limit prolonged exposure to sensitive materials and rotated between different tasks. Institute-provided counseling services were available to the research team throughout the project.} Our study aims to benefit and inform stakeholders to reduce future safety harms for online users.

\subsection{Limitations} 
\label{sec:limitation}

Our study has several limitations. First, our dataset is drawn exclusively from RedNote, a single Chinese social media platform. While RedNote provides rich and diverse narratives of scam-driven human trafficking, these accounts may not represent experiences discussed on other platforms or offline contexts. Relatedly, our keyword-based crawling approach\rev{, which is limited by RedNote's search interface as well as content policy, may lead to bias in the post distribution. Thus, we encourage future work to cross-compare multiple platforms and derive a more comprehensive understanding of the online ecosystems.} \rev{Our filtering combined automated classification with manual validation. While the model achieves a good accuracy, automated classification may have excluded some relevant posts that refer to incidents of scam-driven human trafficking but contain too little textual information to assess their relevance. Additionally, our analysis focuses on textual elements of safety discourse, and we encourage future work to examine multimodal data such as images and videos, which may shed light on different communication strategies and offer complementary insights.} \rev{Last,} our qualitative analysis centers on self-reported narratives shared by those who voluntarily post on social media, and these voices may differ from those that remain silent due to barriers, including trauma, memory lapses, or self-preservation.

\section{Findings}
\label{sec:findings}

\begin{figure*}[t]
  \centering
  \includegraphics[width=\linewidth]{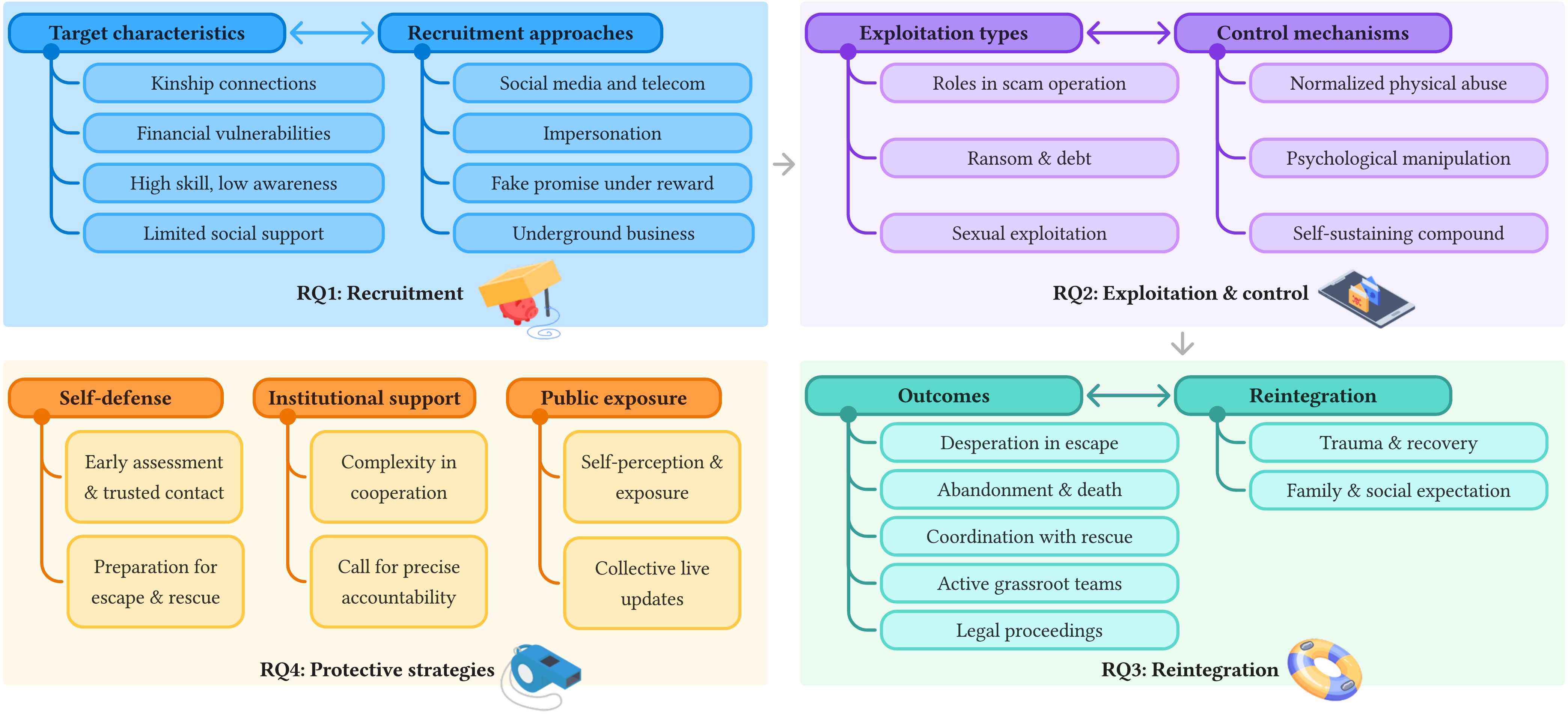}
  \caption{Overview of our findings mapped to scam-to-trafficking phases and research questions: RQ1 characterizes recruitment targets and approaches, RQ2 summarizes exploitation and control mechanisms, RQ3 captures post-trafficking outcomes and reintegration barriers, and RQ4 outlines protective strategies, including self-defense, institutional support, and public exposure.}
  \Description{Four-panel overview diagram mapped to scam-to-trafficking phases and research questions. Top-left (RQ1: Recruitment) summarizes recruitment-related target characteristics (kinship connections, financial vulnerabilities, high skill with low awareness, limited social support) and recruitment approaches (social media/telecom contact, impersonation, fake promises with upfront rewards, underground businesses). Top-right (RQ2: Exploitation and control) summarizes exploitation types (roles in scam operations, ransom and debt exploitation, sexual exploitation) and control mechanisms (normalized physical abuse, psychological manipulation, confinement in self-sustaining compounds). Bottom-right (RQ3: Post-trafficking and reintegration) summarizes outcomes and reintegration barriers (desperation during escape, abandonment, rescue coordination, grassroots rescue advocates, legal proceedings; and reintegration themes of trauma and recovery and family/social expectations). Bottom-left (RQ4: Protective strategies) groups prevention and response into self-defense (early assessment and trusted contacts; preparation for escape and rescue), institutional support (complexity in cooperation; calls for clear accountability), and public exposure (self-disclosure and exposure; collective live updates). Arrows show a top-to-bottom flow from recruitment to exploitation \& control to reintegration, with protective strategies shown in a separate panel of individual, institutional, and public responses.}
  \label{fig:finding_overview}
\end{figure*}

Our findings provide a comprehensive understanding of how scam-driven trafficking is understood and discussed on social media in China. We first report findings around recruitment tactics for scam-driven cross-border human trafficking recognized by RedNote users in \S\ref{sec:RQ1}. We then describe results from our analysis on the exploitation and control mechanisms shared by RedNote users in \S\ref{sec:RQ2}. Then, findings about the post-trafficking outcomes and reintegration challenges identified by users affected by scam-driven trafficking are presented in \S\ref{sec:RQ3}. Finally, we show how RedNote users share and evaluate protective strategies against scam-driven human trafficking in \S\ref{sec:RQ4}. Figure~\ref{fig:finding_overview} shows an overview of our findings.

\subsection{RQ1: Recruitment}
\label{sec:RQ1}

In this section, we discuss the recruitment tactics for scam-driven cross-border trafficking as recognized and discussed by RedNote users. RedNote users have identified a range of targeted tactics, leveraging people's differing cultural, financial, and educational backgrounds in China (\S\ref{sec:target_characteristics}) when recruiting or defrauding people for scam-driven human trafficking (\S\ref{sec:recruitment_approaches}).

\subsubsection{Target characteristics.}
\label{sec:target_characteristics}

RedNote posts collectively reveal four key target characteristics that perpetrators exploit. Our analysis showed that recruitment scams consistently leverage people's financial vulnerabilities and traditional Chinese kinship culture. Perpetrators specifically look for potential victims who possess skills for scam operations while having limited protection against exploitation.

\paragraph{\textbf{Kinship and cultural connections.}} In multiple cases, perpetrators exploited family ties and cultural connections when approaching initial targets and abusing trust. \texttt{P-e8Y} gave an example of a recruitment scam spread through family networks. The target carried the thought that \textit{``there is no way my [her] own sister would scam me [the target]''}, who also received \textyen50,000 upfront from her sister, luring her, and eventually several other senior family members, to migrate to Cambodia for fraud working.

Community interpretations on RedNote suggest that such exploitation operates through the unique kinship ties in China~\cite{chun-ca96} where shared local identities, family and clan reputations, and lineage-based networks provide falsely justified legitimacy and recruitment opportunities. Multiple RedNote posts point out that the perpetrators and victims are from the Fujian Province, which is a major hub of overseas migration from China and is associated with one of the largest online scam networks globally~\cite{theguardian-2024}. For example, in \texttt{P-Nh4}, a proclaimed insider commented that \textit{``almost all managers are Fujianese''} for a scam compound in the Shan State, Myanmar. Another poster condemned these perpetrators and thought they \textit{``brought massive negative impacts to the clan and the society and would possibly deserve expulsion from their pedigrees''}, which further showcases how the traditional kinship culture shapes people's understanding of modern cybercrime (\texttt{P-Jo7}).

Beyond initial recruitment, \texttt{P-U0v} discusses that victims' kinship ties can be further unwillingly exploited during forced recruitment scam production, as the perpetrators will be \textit{``squeezing the last bit of use from victims and forcing them to proclaim to relatives and friends their infinite success abroad''} to lure more victims to scam compounds. This example further reflects the broader social codes in China, i.e., the so-called guanxi~\cite{chen2004intricacies}, where perpetrators exploit based on reciprocal obligations, honor, and mutual trust, extending the recruitment channel from close family members to others in the network, e.g., \textit{``camaraderie of a friend''} (\texttt{P-5mn}, \texttt{P-B9j}).

\paragraph{\rev{\textbf{Individual and generational financial vulnerabilities.}}}
Perpetrators consistently exploit victims' aspirations for financial advancement by crafting deceptive employment opportunities that appear to align with personal backgrounds and career interests. One case describes a mother's account of her son, who became disillusioned with unstable factory employment and the desire to pay medical bills for her. The son was subsequently targeted by a \textit{``game company''} in Yunnan, resonating her son's interest in e-commerce and \textit{``professional background''} in gaming, with a monthly salary of over ten thousand Chinese Yuan (\textyen), but ended up trafficking her son to Myanmar (\texttt{P-H7j}). Multiple other survivors shared the same financial hardship or the desire to \textit{``make a big fortune''} (\texttt{P-w81}). Some posts questioned if the described financial vulnerabilities and aspirations are simply greed, describing survivors as \textit{``freeloaders''} (\texttt{P-N8k}). \add{Beyond individual vulnerability, RedNote users connected individual financial vulnerabilities to generational economic crisis, revealing their concerns in social safety nets and labor market instability. For instance, they called for \textit{``strengthening protection of workers' rights''} (\texttt{P-v68}) for removing the financial driver of being lured into scam-driven human trafficking. They also recognized the inefficacy of policing strategies such as border control because \textit{``Yunnan border guards can't stop people who think they can make money''} (\texttt{P-p6a}).}

\paragraph{\textbf{High skill potential but low awareness.}} Traffickers deliberately target individuals from a mix of educational and age backgrounds who possess the skills or potential to be converted into scam workers. The targets perpetrators look for include multi-lingual people (\texttt{P-A6z}) to reach prospective victims internationally. Targets also include teenagers and college students who are digitally literate, e.g., with social media platforms such as Kuaishou, who can be trained quickly for fraud production. These young targets are often depicted as lacking adequate social experience, legal understanding, and geographical awareness against scams, especially during travels abroad, which is illustrated by \textit{``followed online acquaintances from Guangdong to Guangxi and then across borders, only realizing belatedly that they were being trafficked''} (\texttt{P-J1v}).

\paragraph{\textbf{Limited social protection and support.}} Perpetrators target populations with high mobility to travel but limited social protections at home and abroad. These targets include \textit{``left-behind children,''} a large population of Chinese children who stay in rural hometowns while their parents work in urban areas~\cite{jingzhong2011left} (\texttt{P-J1v}), as well as tourists who \textit{``had no idea why they are sent to borders between Thailand and Myanmar''} or got trafficked by the \textit{``travel partners''} they just met (\texttt{P-5yt}). Additionally, RedNote users also offered travel warnings specific to women, e.g., \textit{``especially single and beautiful girls,''} who might lack the physical strength to fight against street-kidnapping for sexual exploitation in the scam compounds (\texttt{P-1VH}). The post also commented that \textit{``no one will look after you''} during their travels.

\subsubsection{Recruitment approaches}
\label{sec:recruitment_approaches}

Perpetrators often combine multiple tactics to create more convincing deceptions that make it harder for potential victims to recognize the danger.

\paragraph{\textbf{\rev{Communication transiting targets across multiple channels exploiting digital governance blindspots.}}} Social media arised as the dominant channel for perpetrators to broadly disseminate fraudulent job advertisements and romantic interests, and to gradually cultivate trust of the target in the perpetrator. These social media platforms include apps that are not officially allowed in China, for example, Facebook and Telegram, which are often referred to in coded language such as \textit{``面子书''} (``face book'' -- Facebook) and \textit{``飞机群''} (``airplane group'' -- Telegram) (\texttt{P-3BQ} and \texttt{P-t3u}). Additionally, posters discussed seeing recruitment posts on several domestic social media apps in China other than RedNote, such as Douyin and WeChat (\texttt{P-z3p} and \texttt{P-v3S}). Telephone calls are used for more direct and urgent communication, and when used along with other social media channels, they create a veneer of formality for business-related recruitment scams, e.g., directly with the \textit{``boss''} (\texttt{P-3BQ}). \add{One user described how \textit{``initial contact through chat on Douyin, creates romantic emotional bonding and then introduces fake investment websites''} (\texttt{P-z3p}). Posts reveal that perpetrators strategically choose platforms with weaker governance mechanisms. For instance, one poster warned others (\texttt{P-1Jy}) by saying \textit{``they generally use chat software that Chinese people don't commonly use, because there's no strict control mechanism.''} This pattern shows that recruitment shifts to foreign platforms with less oversight. When communication moves across platforms, users receive no indication that they are crossing into spaces with weaker protections, allowing the trust built on regulated platforms to carry over without recognizing the risk.}

\paragraph{\textbf{Sophistically organized impersonation.}} Perpetrators impersonate distant relatives, authority figures, corporate representatives, or romantic partners when luring potential targets. Such impersonation can be sophistically organized, e.g., posing as clients to approach freelance workers such as dancers, magicians, makeup artists, and photographers, offering overseas work opportunities with \textit{``accommodation and flights provided''} through seemingly normal business inquiries. More specifically, a news report indicated that perpetrators would run a seemingly legitimate business and later traffic targets to Cambodia for scam work \textit{``during a corporate retreat''}, involving a group of impersonators (\texttt{P-5mn}).

\paragraph{\textbf{Fake promises under upfront rewards.}} We observed that perpetrators are willing to pay for an upfront reward to lure targets and leverage their financial vulnerabilities, especially in fraudulent recruitment and investment opportunities. These upfront rewards include passport renewal fees (\texttt{P-3BQ}), free travel and transportation (\texttt{P-B9j}), advanced training payment (e.g., \textit{``gave her 50k to receive manicure training''}), and express loan or startup budget (\texttt{P-e8Y}). These apparent benefits convince potential victims that they are entering profitable arrangements. However, survivors recounted how these promises are later transformed into mechanisms of entrapment, with upfront rewards reinterpreted as debts and leveraged for control once victims arrive at scam compounds.

\paragraph{\textbf{Willful participation in underground business.}} Our analysis reveals that trafficking recruitment deliberately exploits the boundaries between legitimate high-paying opportunities and underground business in China and Southeast Asia. Posts reveal several smuggling schemes leveraged by perpetrators to attract people (\textit{``smuggling gold''}), courier services for high-value, export-controlled goods (\textit{``transporting edible bird's nests''}~\cite{shukri2018consumer} for \textit{``\textyen15k per trip''}), or other semi-legitimate cross-border trade that promises exceptional returns (\texttt{P-E8q}). This exploits victims' intentions for \textit{``quick money''} to be tempted, as they may see this work involve bending rules and would not constitute serious criminality. Another underground business mentioned is the \textit{``debt trap,''}  and a user suspected that some targets may have intended to \textit{``get something for nothing''} and never intended to repay the loans. In particular, several posts explicitly accused survivors' informed participation in scam work, who \textit{``went specifically for rebate fraud''} (\texttt{P-SI6}). The same post also claimed that a large proportion of \textit{``about 90\% are not real victims but participate in overseas fraud activities.''} Similarly, \texttt{P-9og} openly questioned other survivors' proclaimed innocence by saying that \textit{``he dares to write, and you dare to believe it.''}

\subsection{RQ2: Exploitation and Control}
\label{sec:RQ2}

We observed that RedNote users share accounts of diverse tactics that traffickers employ to exploit and control victims. 

\subsubsection{Exploitation Types}
\label{sec:exploitation_types} 

Trafficked victims are exploited for their physical and intellectual labor in scam production, alongside additional financial and sexual values.

\paragraph{\textbf{Trafficked victims transformed into specialized roles within scam operations.}} Our analysis reveals that survivors are transformed into scam workers in specialized organizational roles, along with the perpetrators. Labor is structured around scam production quotas, with survivors assigned to various operations including phone scams and chat-based fraud, social media farming, and romance grooming. One survivor described being \textit{``required to add 5 customers (scam targets) online every day''} (\texttt{P-Q1T}). Training in using specific scam tactics is given when new ``fraud workers'' [survivors] receive training with scripts and materials: \textit{``Given a thick stack of conversation materials to memorize, plus twenty phones to play various identities''} (\texttt{P-Gy2}). One victim documented that they got trained for a multi-stage romance scam targeting women aged 35-60 with a predetermined geographic area and timeline: \textit{``The scam uses a military veteran's persona and gaming to build trust, then foster fake romantic ties of `husband and wife' before pushing a 4-day money making cycle. Avoiding remote regions like Yunnan and Guizhou''} (\texttt{P-z3p}). The exploitation of young victims is particularly striking, with one observer noting (\texttt{P-Q1T}) \textit{``I've seen the youngest dog pusher and agent at 19 years old, using a childish voice to impersonate a nearly 40-year-old middle-aged man,''} demonstrating how traffickers force even teenager victims to adopt false personas far beyond their years. \rev{Victims may eventually get ``promoted'' to management roles within scam compounds, such as enforcers of disciplinary actions or coordinators for scam operations. The term \textit{``Piglets (猪仔)''} is used by traffickers and within the online community to refer to trafficked victims, framing them as} livestock within a trading system for labor exploitation between different scam operations and compounds. Victims reported inflation of their \textit{``buy-out cost''} upon transfer between scam compounds, with documented cases showing \textit{``my buy-out cost increasing from \$1,300 to \$13,000''} (\texttt{P-HF3}). This commodification extends to the most extreme forms of exploitation, with survivors reporting witnessing \textit{``organ harvesting''} and describing the compounds as places where victims face the ultimate threat of bodily dismemberment when they become unprofitable or uncooperative (\texttt{P-x9a} and \texttt{P-5Ig}).

\paragraph{\textbf{Ransom demands and debt manipulation targeting victims' families.}} \rev{Exploitation extends beyond forced labor and commodification of the victim to their families through ransom demands and debt manipulation, described as \textit{``squeezing the last bit of use from victims''} by one survivor (\texttt{P-U0v}). Ransom videos showing tortured and pleas of victims are sent to families, pushing already impoverished families deeper into debt, with families lamenting \textit{``Never mind 300,000 [RMB], our rural family can't even come up with 30,000.''} In several posts (e.g., \texttt{P-Q9F}), victims initially pursued higher wages to cover family medical bills, yet their families ultimately sold assets and borrowed heavily to make up the ransom amount, revealing a vicious cycle where economic desperation leads to victimization, and victimization creates even greater economic desperation. Some posts also reference established price ranges, such as \textit{``Ransom is often between 200,000 and 400,000 RMB.''}} \add{Ransom demands also create uncertainty and moral pressure. Relatives worry that pleas for help may be secondary scams but fear the cost of ignoring them. Many turn to RedNote to crowdsource verification and advice. One user created \textit{``an alt account''} for anonymity and wrote: \textit{``several friends say it's a scam, but it's hard to determine when it involves someone's life---if it's not a scam, my little cousin might get trapped, and I don't know whether to lend this [money emoji]'')} (\texttt{P-GW6}). Another one asked, \textit{``My relatives were scammed in the Myanmar Industrial Park! I see this kind of help request every now and then. Can they really be saved?''} (\texttt{P-i3h}). These narratives show families using the platform to narrate financial and emotional strain and to collectively reason about whether, and how to respond to ransom demands.}

\paragraph{\textbf{Sex exploitation in parallel with scam operations.}} Our findings reveal that female victims face additional exploitation within scam-driven trafficking. While initially forced into fraud work, women face sexual exploitation threats when resisting productivity demands, whereas males face physical punishments like beatings and electric shocks. Male perpetrators leverage romantic relationships and power imbalance for sexual exploitation and control. One post (\texttt{P-Gy2}) described a case where a scam operator \textit{``has more than ten girls under his command,''} who \textit{``often pretends to be tall, rich and handsome online to gain the trust of girls, and then takes them to Myawaddy, Myanmar.''} Physical attributes become commodification criteria that determine the exploitation pathway, where \textit{``slightly prettier girls will be targeted by supervisors, directors, and team leaders''} (\texttt{P-xL7}). Those who refuse fraud work or fail to meet performance metrics face immediate transfer to \textit{``clubs,''} where phone access is eliminated entirely, severing their final communication lifeline (\texttt{P-xL7}). The term \textit{``handrail girl (扶手女)''} is a localized label that refers to women who, after being drugged or coerced, were forced into auxiliary sexual roles in brothels.

\subsubsection{Control Mechanisms}
\label{sec:control_mechanisms}

To exploit victims in a ``sustainable'' manner, perpetrators apply a range of physical and psychological control.

\paragraph{\textbf{Normalized physical abuse.}} Violence functions as a normalized productivity enforcement tool explicitly tied to fraud quotas and rule compliance within scam-driven trafficking operations. Multiple first-person accounts described punishment regimes where survivors face systematic physical abuse for failing to meet performance targets, with one post stating (\texttt{P-0gV}) \textit{``if you miss targets you are shocked and beaten.''} A post by a self-proclaimed victim in diary-like format documented mass punishment and shaming exercises involving \textit{``frog jump''} drills with \textit{``over a hundred people''} and brutal punishment of two young escapees (aged 14 and 17), with the 14-year-old electrocuted on the forehead and ear \textit{``for nearly a minute''} before confinement in a coffin-sized \textit{``black room.''} Seemingly innocent phrases like \textit{``eating popsicles''} and \textit{``drinking bubble tea''} mask descriptions of brutal torture methods involving dismemberment and violence (\texttt{P-5Ig}).

The psychological impact manifests through behavioral conditioning, as evidenced by a 17-year-old victim who \textit{``developed the habit of immediately lying face down with buttocks raised in preparation for beating upon waking each morning,''} demonstrating how physical control transforms victims from resistance to resignation (\textit{``Since I'm already here, I might as well just do the job'')} (\texttt{P-0gV}). Prolonged work hours requiring 700--800 scam calls per day create physical and psychological burnout, pushing victims from initial resistance toward resigned compliance.

\paragraph{\textbf{Psychological manipulation by exploiting cultural values and drugs.}} Our analysis reveals how perpetrators exploit cultural values and create dependency to maintain victim compliance. As one pro-claimed survivor (\texttt{P-uY1}) states: \textit{``She deceived all the agents by saying they were her younger brothers, saying how could her sister treat you badly \dots carefully look at those who follow her, one by one either taking drugs or becoming inhuman and ghostlike, or mentally unclear.''} This false familial manipulation of perpetrators positioning themselves as elder siblings exploits the Chinese Confucian principle of filial piety and fraternal duty~\cite{pei2024confucian} that positions older siblings as figure deserving respect and obedience from younger family members, making resistance feel like a betrayal of core cultural values. This also exploits victims' need for protection in isolated environments while undermining their autonomy through drug dependency and psychological destabilization. Posts showed that victims eventually exhibit \textit{``willingness to give money to traffickers''} (\texttt{P-th4}) and resist rescue attempts after prolonged conditioning. Perpetrators additionally use drugs to enhance psychological manipulation and ensure compliance, through forced consumption of \textit{``happy water with drugs,''} \textit{``drugged water,''} or Ketamine (\texttt{P-w9M}).

\paragraph{\textbf{Self-sustained compound facility for isolation and control.}} \rev{Our analysis further reveals survivors or insiders describing how the scam compound facilities enable isolation and control. Some survivors described the scam compound as \textit{``prison-like''} with constant surveillance and degrading conditions, such as \textit{``sleeping on bare wooden boards \dots with cameras in the toilets and showers under 24/7 surveillance, leaving us no privacy at all''} (\texttt{P-f9O}). The long-term effect is a \textit{``zombified''} workforce that has little energy left to resist. In contrast to the harsh work conditions, some users depicted compounds as self-contained cities with victims describing \textit{``Compounds features casinos, KTV, entertainment centers, banks, supermarkets, everything is available''} (\texttt{P-Gy2}),} which allows victims to slowly become accustomed to their captivity. One post provided an insider perspective on their lives as a contracted scam worker: \textit{``the bonus I received was not much. I earned money from the compound and spent it in the compound, leaving nothing to take home''} (\texttt{P-K2M}).

\subsection{RQ3: Post-Trafficking and Reintegration}
\label{sec:RQ3}

Below we analyze the barriers people face when escaping from the scam compounds and reintegrating into their normal lives. 

\paragraph{Post-Trafficking Outcomes}
\label{sec:post_trafficking_outcomes}
Our analysis reveals a complex  landscape of post-trafficking outcomes, ranging from successful escape and rescue to abandonment, re-trafficking, and death. For those who return, the path to reintegration is fraught with significant medical, legal, and social challenges.

\paragraph{\textbf{Desperation in escape.}} Escape attempts represent an act of desperation and victims' final chance for survival. Some victims resort to severe self-harm, as documented by an awareness spreader: \textit{``some are willing to break their hands and feet to jump from buildings to escape''} (\texttt{P-7TG}), while others sent desperate social media pleas for help: \textit{``This is my only chance to send a distress signal, perhaps also the last time, please save us''} (\texttt{P-J3T}). However, even successful escapes often lead to recapture as one escapee  \textit{``was sold by locals to local armed forces,''} while recaptured face even more brutal punishment as a deliberate deterrent, including being \textit{``beaten unconscious multiple times per day and having boiling water poured on them''} (\texttt{P-C1C}).

\paragraph{\textbf{Abandonment, being re-trafficked and death under extreme pressure.}} Our analysis reveals how perpetrators dispose of victims who become unprofitable or lack useful skills. One post stated, \textit{``If you're illiterate or can't type, they'll just beat you up and dump you on the roadside to fend for yourself''} (\texttt{P-K2M}). However, escape or release does not guarantee freedom, as many survivors face re-trafficking to other compounds: \textit{``I was sold by human traffickers to the Myawaddy fraud park, where I was sold more than ten times''} (\texttt{P-E9y}). Others documented the extreme working conditions of \textit{``17--18 hours of forced work per day with corporal punishment''} that left victims \textit{``dying like slaves''} (\texttt{P-J3T}), while others die from \textit{``exhausted to death by the high pressure''} (\texttt{P-7aC}), overwork, or \textit{``Beaten to death, tortured to death, and even committed suicide''} (\texttt{P-C1C}).

\paragraph{\textbf{Coordination with institutional rescue and public support.}} RedNote users documented how institutional rescue efforts achieved notable success through international coordination, yet face significant barriers. Interventions often involve coordinated police, where \textit{``Thai authorities are coordinating with the Chinese embassy in Thailand''} (\texttt{P-v68}), diplomatic channels, and family advocacy. Successful cases often involve multi-national cooperation, as documented when \textit{``Thai police successfully arrested one of the suspects... Thai authorities are coordinating with the Chinese Embassy in Thailand to send the victim back to China''} (\texttt{P-v68}). 

Several families shared detailed information, including departure timeline and locations, with some publicly expressing willingness to pay for ransom (\texttt{P-9zh}): \textit{``younger sister was trafficked to Myawaddy on September 15...I hope the company that has her won't beat or scold her. Please just confirm your safety; compensation is negotiable.''} However, failures may occur due to victim non-cooperation, as in one case where, despite \textit{``cooperation of police from two countries, [they] finally found the child at Myanmar customs, but he seemed to have been brainwashed... unwilling to contact parents and friends''} (\texttt{P-th4}), demonstrating how psychological manipulation can undermine successful interventions. Furthermore, rescue effectiveness varies significantly by survivors' choice, with many opting for personal networks (through WeChat) over authorities. This is evidenced in a news report where victims \textit{``did not choose to call police''} but rather \textit{``sought help from friends through WeChat.''} Reports indicate that survivors who escaped suffer serious psychological damage and distrust outsiders, with some choosing silence over reporting due to death threats from perpetrators and concerns about collusion in the rescue teams (\texttt{P-Zj0}).

\paragraph{\textbf{Active presence of grassroots rescue teams.}} Beyond formal law enforcement channels, our analysis highlights the emergence of grassroots rescue networks operating through RedNote. These self-proclaimed \textit{``anti-trafficking rescue teams''} actively advertise service timelines such as \textit{``response times as fast as `20 minutes' for driver deployment''} and \textit{``will act within 48 hours and rescue within 3 days''} (\texttt{P-H7j}, \texttt{P-UU2}). To further establish credibility, some agree to authenticate via trusted intermediaries, e.g., \textit{``Cap Uncle (帽子叔叔, slang referring to the police)''} (\texttt{P-F2H}, \texttt{P-9zh}, and \texttt{P-e9F}) as well as signing formal \textit{``service agreements''} (\texttt{P-1nS}) with families. These teams offer detailed accounts of past successful rescue cases and rescue procedures, including border-crossing logistics and administrative processes; for example, descriptions of Thai immigration procedures such as \textit{``Mae Sot Immigration Bureau, detention 20--30 days, 55,000 baht for expedited processing'' } suggest insider or experiential knowledge (\texttt{P-3jU}). They actively post advice by warning against contacting local police in compromised jurisdictions and warn \textit{``locals get \$3000 rewards for returning escaped victims to compounds''} (\texttt{P-m6R}, \texttt{P-D2Y}) and instead recommend hiding first, then coordinated outreach to embassies and vetted rescue teams.

Family responses to grassroots teams are mixed. While some treated them as vital lifelines, others rejected their help outright. In one case (\texttt{P-UU2}), a volunteer recounted contacting a mother about her son in Cambodia who refused to provide help because they viewed the victim as a burden or as unfilial, saying he had \textit{``spent a lot of the family's resources [money]... and thrown the household into complete disorder,''} and concluded, \textit{``there is no more money left to save him, so we will just stop caring about him.''} Finally, not all grassroots rescues succeed. Failed rescues were thwarted by armed perpetrators, with the rescue team saying  \textit{``We did all we could... unless something unexpected happened, the person is already [back] inside the mountain compound''} (\texttt{P-yY6}).

\paragraph{\textbf{Legal proceedings.}} RedNote posts show how official narratives frame (voluntary surrender) as the only path to leniency under an expanding legal crackdown. Users actively relayed such warnings that \textit{``from May 1 onwards, all individuals engaged in overseas telecom fraud---whether voluntary or coerced---will be punished under the new law. Only those who surrender to public security may receive mitigation or exemption''} (\texttt{P-3fm}). These announcements were reinforced by posts citing state media reports that \textit{``44,000 suspects have been transferred back, including 171 ringleaders''} (\texttt{P-x9a}). The user also recounted stories of returnees who, after being rescued through family payments, later chose to turn themselves in out of conscience. \add{These accounts suggest that survivors and families view the legal environment as highly punitive and are unsure whether coerced participants will be treated as offenders or as victims, shaping their decisions about escape, reporting, and surrender.}

\subsubsection{Reintegration}
\label{sec:reintegration}

Survivors experience trauma when recovering from trafficking experiences, however, some still face family and social pressure.

\paragraph{\textbf{Trauma and prolonged recovery.}} Our analysis of reintegration challenges reveals how trafficking survivors face long-term complex recovery. Medical supervision becomes necessary for many victims, as documented in one case where a 17-year-old survivor \textit{``had been tortured to mental breakdown, even his basic walking posture was abnormal \dots Every morning when he woke up, he would lie down with his buttocks raised in preparation for being beaten''} (\texttt{P-YP8}). This example showcases how trafficking trauma manifests through conditioned behavioral responses that require extensive psychological intervention. Compulsory rehabilitation becomes necessary for victims forced into drug dependency during captivity. One account describes how a victim was \textit{``forced to consume drugs `K powder' \dots She will face compulsory drug rehabilitation and psychiatric treatment''} (\texttt{P-c0o}), where the control mechanisms imposed during trafficking create long-term barriers to recovery.

\paragraph{\textbf{Family roles in reintegration and social expectation.}} Survivors who returned home may face intensive and compulsory treatments. In one case (\texttt{P-c0o}), the survivor required compulsory drug rehabilitation and psychiatric care after months of violence and forced drug use in Cambodia. Recovery depended heavily on \textit{``requires her family's ongoing care and support.''} Some families implemented strict surveillance of rescued victims as they are \textit{``constantly watched by parents, confined at home and not allowed to go out''}  (\texttt{P-th4}), reflecting deep mistrust after what the victim has been through. Beyond immediate family dynamics, survivors also needed to deal with judgment from the wider social circles and online communities. As one user commented on a successful rescue (\texttt{P-wy1}): \textit{``I hope she learns her lesson and goes on to live smoothly in the future,''} reintegration is often framed not only as recovery but also as a moral correction.

\subsection{RQ4: Protective Strategies}
\label{sec:RQ4}

RedNote users have developed protective strategies that emerge from collective trauma and transformed individual experiences into community wisdom. Our analysis identifies four main themes in users' proposed protective strategies, ranging from self-defense to navigating institutional responses and support.

\subsubsection{Self-defense.}
\label{sec:self_defense}

RedNote users have recognized several opportunities and the critical intervention windows to defend themselves and avoid subsequent harms, especially in the early stages of scamming and recruitment.

\paragraph{\textbf{\rev{Early assessment and trusted contact for different demographics.}}} RedNote users offered tips for people from different backgrounds to self-assess the risks in a potential scam. For individuals seeking overseas employment opportunities, users noticed several red flags in job ads, including \textit{``high pay for minimal work,''} \textit{``all-expenses-paid round-trip tickets,''} \textit{``only reveal location after you arrive,''} and \textit{``boast extensively but won't show you company videos or photos''} (\texttt{P-1Jy}), \textit{``temporary flight adjustments''} (\texttt{P-5XF}), movements through specific high-risk routes that bypasses official checkpoints such as \textit{``Mae Sot and Tak in Thailand toward Myawaddy in Myanmar''} (\texttt{P-3jU}), and requirements to \textit{``bring China-issued bank cards abroad''} (\texttt{P-Y9s}). Rather than suggesting turning down all potential opportunities, users highlighted actionable verification steps, including confirming company legal registration, demanding authentic labor contracts and proper work visas, and rejecting offers requiring \textit{``tourist visas''} or vague \textit{``training/team-building''} arrangements (\texttt{P-5XF}).

\add{When users posted warnings to raise awareness, we observed different strategies depending on the target audience's gender. 
Posts targeting women commonly use hashtags such as \texttt{\#girlsmustsee}, \texttt{\#femalesafety}, and \texttt{\#GirlsTalk}, creating dedicated channels for delivering warnings to  female users (\texttt{P-F2d}).} \add{Users shared} specific suggestions for female users against grooming or romantic scams, which may gradually steer victims towards sex trafficking and forced participation in \textit{``client services''} on scam campuses such as Myawaddy. Protective strategies for women emphasize maintaining communication independence: retaining phone access and location services, as well as avoiding \textit{``solo meetings''} and \textit{``free trips.''} \add{By contrast, warnings involving men rarely employ demographic-specific hashtags, instead relying on generic employment-fraud warnings or highlighting extreme physical torture, forced labor, and organ trade (\texttt{P-OI6}).}

International students are highlighted as another vulnerable group. RedNote users emphasized protective strategies specific to the student context, including maintaining skepticism toward unsolicited contact from strangers, avoiding any money transfer requests regardless of the claimed authority source, and ensuring regular communication with family members. The community particularly warned against schemes involving \textit{``embassy, online shopping, relatives borrowing money, online gambling, online pornography''} (\texttt{P-0gV}) as common tactics used to target international students who may be more vulnerable due to their distance from family support systems. All these early intervention strategies highlight the need to build a reliable support network which may involve family members and friends, with whom one can share detailed itineraries and emergency contacts: \textit{``report your itinerary to family and friends, letting them know your whereabouts''} prior to a job visit (\texttt{P-E3U}); or trusted college staff who can help verify the legitimacy of international financial transactions. Additionally, it is crucial to maintain stable communication channels and regular check-ins, even using \textit{``code words''} such as \textit{``My brother died in northern Myanmar''} which, as noted in \texttt{P-9og}, \textit{``some rescue teams can recognize.''}

\paragraph{\textbf{Self-preparation for escape and rescue.}}
Although the ideal timing to mitigate further human trafficking risks is prior to a foreign travel, RedNote users still identified several opportunities that increase the chances of escaping or being rescued by collecting evidence and establishing communication channels. First, the \textit{``waystations''} such as farms, barracks, or private houses where survivors are detained during transportation present chances for survivors to (\texttt{P-L8z}) \textit{``record vehicle license plates and driver details upon boarding''}. This information may serve as evidence that enables upstream arrests and facilitates negotiated release. Additionally, users also emphasized the practical considerations of installing Google Maps for location tracking as \textit{``domestic Chinese navigation services become unreliable abroad''} (\texttt{P-Y9s}).

For individuals already trapped within trafficking compounds, RedNote users provided detailed intelligence-gathering protocols to facilitate rescue operations. Critical information includes \textit{``compound names, company operational codes, specific building and room numbers, identification documentation, precise geolocation coordinates, and recent photographs''} (\texttt{P-H1K}). Users also documented creative escape strategies from successful escape experiences, including exploiting \textit{``medical emergencies excuse''} (\texttt{P-T1T}) as opportunities to leave compounds under supervision. However, community guidance strongly advises against pre-payment and recommends insisting on \textit{``simultaneous person-for-payment exchanges''} when negotiation becomes unavoidable. This reflects hard-learned lessons from cases where advance payments resulted in continued captivity or transfer to additional criminal networks.

\subsubsection{Institutional support.}  
\label{sec:institutional_support}

Although RedNote users recognized the complication in receiving institutional support internationally, they expressed their expectation for more precise accountability measures.

\paragraph{\textbf{Complexity in cross-border cooperation.}}
RedNote users recounted experiences with multiple institutional stakeholders in rescues, highlighting both their contributions and the difficulties of cross-border coordination. Chinese embassies play a key role in victim verification and arranging rescues, often credited with \textit{``coordinating with local police to take my brother safely across the border''} (\texttt{P-L1M}), or with Thai immigration police in joint rescues of Chinese victims (\texttt{P-v3S}). \texttt{P-U6m} demonstrates how families were advised to \textit{``contact the embassy immediately --- embassy staff will arrange personnel to help find and confirm victim safety.''} Embassy personnel coordinate complex rescue operations across multiple jurisdictions, as described in one case where embassy staff \textit{``arranged a contact to visit my brother [victim] and allowed me to video call him, confirming their location was safe''} (\texttt{P-U6m}). However, the diplomatic process requires navigating complex international relationships and local corruption challenges.

Thai immigration authorities manage detention and processing systems that differentiate based on victims' documentation status. \texttt{P-3jU} details how victims with valid passports are \textit{``transferred to Thailand's Mae Sot Immigration Bureau, detained 20--30 days (can pay 55,000 baht for expedited processing), then transferred to the Bangkok Immigration Bureau''} and extended detention periods for victims without valid passports. The immigration system offers expedited processing options, but at substantial financial cost, including \textit{``1,200 baht for family contact, ticket money, prison hygiene fees, escort fees, and expedited fees (20,000 baht).''} Police agencies provide the formal reporting mechanism and coordinate with international partners. However, their effectiveness varies significantly across jurisdictions. A victim's relative recounted that domestic police often have limited authority in cross-border cases, with officers stating \textit{``we have no law enforcement or investigative authority''} and advised to contact the Wa State police in Myanmar (\texttt{P-u9F}). The Wa State police then required a formal request letter from the Chinese public security authorities before taking any action. Families are required to navigate multiple institutions with overlapping but constrained responsibilities.

\paragraph{\textbf{Calls for more precise accountability measures.}}
User attitudes reveal growing frustrations with genuine victim authenticity. \texttt{P-Z2A} claims that \textit{``about 90\% are not real victims but participate in overseas fraud activities.''} This skepticism has led to calls for preventative accountability measures, with users demanding \textit{``passport blacklisting for rescued survivors for at least three years as a lesson to prevent recurrence''} (\texttt{P-Z2A}). Such proposals reflect the community's desire for stricter enforcement mechanisms that would deter voluntary participation in overseas fraud operations. Institutional support relies on coordination between embassies, police, immigration authorities, and legal advocates, each offering essential but limited services within their jurisdictions. While these institutions enable successful rescues and legal protections, the multi-stakeholder system creates bureaucratic delays and mixed rescue outcomes. At the same time, community debates over victim authenticity advocate for stronger preventative measures. Nevertheless, \texttt{P-VL5} illustrates a contrasting view from a defense lawyer that many survivors should be \textit{``treated with leniency''} in lawsuits, because they \textit{``had no malicious intent, and did no substantial work there.''} They also noted unclarities and challenges in such lawsuits as \textit{``many witnesses' testimonies were fake.''}

\subsubsection{Public exposure.}
\label{sec:public_exposure}

We find that RedNote serves as a channel for people who have experienced scam-driven human trafficking, including self-proclaimed survivors or scam workers, to raise public awareness by exposing details in the ecosystem.

\paragraph{\textbf{\rev{Survivor self-perception and exposure.}}}
\rev{Self-identified survivors revealed contested understandings of their roles in scam operations. Their self-perception is shaped by both evolving awareness through engagement in scam work and public moral judgment. Many initially viewed their work as legitimate employment and only realized the deception later. One post recounted that \textit{``only then did she realize that the so-called high-paying job was actually a telecom fraud scheme''} (\texttt{P-Gy2}). Others knew from the outset that they were entering illicit work. One described themselves as a \textit{``volunteer for this scam''} (\texttt{P-uY1}). To warn others, survivors strategically exposed recruitment tactics by sharing scammers' private communications. \texttt{P-BN3} includes messages from \textit{``a Cambodian scam company boss who wanted me to help recruit.''} They positioned their testimony as cautionary advice to \textit{``make those who want to pursue gold-digging or high-salary jobs abroad give up their dreams.''} Under violence and surveillance, survivors adopted survival-focused mindsets. One stated \textit{``Since I'm already here, I might as well just do the job''} (\texttt{P-0gV}) while observing coworkers becoming \textit{``inhuman and ghostlike''} (\texttt{P-uY1}). At the same time, survivors recognized they are participating in fraud. One admitted \textit{``If I say that those of us who depend on this industry are all not good people, I really have no way to refute that''} (\texttt{P-uY1}). Their voices include those who were deceived and now \textit{``hate [themselves] for being too naive.''} They also include those who \textit{``voluntarily participated in fraud.''}} 

Public reactions toward survivors' disclosure alternate between sympathy and condemnation. Some view survivors as purely coerced, while others view them as greedy or complicit. Some insisted \textit{``No one can force you [survivors] to go if you don't want to.''} Posts invoke \textit{``perfect victim''} expectations and debate \textit{``whether they [survivors] should be condemned.''} This judgment produces self-censorship among survivors. One testimony preemptively requests \textit{``hoping for no malicious criticism''} (\texttt{P-3BQ}). Survivors' self-understanding thus remains unstable. 
Altogether, survivors also view themselves as a hybrid of victims of coercion and perpetrators of fraud in response to public judgment.

\paragraph{\textbf{Collective live updates.}}
Victims, survivors, and rescue teams leverage RedNote to share live updates regarding individual status and the progress of rescue campaigns, extending beyond simple awareness posts. Victims trapped in scam compounds provide ongoing status reports about their situations to maintain a connection with the outside world and create digital evidence trails of their experiences. In one example, one self-identified victim said (\texttt{P-tK4}): \textit{``if I don't post updates for a long while, that means I am close to death.''} Beyond personal survival communications, victims and insiders provide detailed updates about compound activities such as enforcement deadlines, organizational changes, victim transfer patterns, punishment and location of execution. For example, a post shared that a local armed force will \textit{``hand over a batch of around 3,000 dog pushers back to China through the Chinese and Thai cap uncles [the police] this year.''} These updates serve both as warnings and as informal reporting channels, which enable families and rescue groups to verify compound activities, locate victims, and coordinate interventions. Taken together, RedNote users transform individual experiences into collective protective strategies, ranging from self-defense and escape preparation to survivor testimony and collective live updates. These practices strengthen vigilance, expand support networks, and serve as cautionary signals that help others avoid victimization.

\section{Discussion}
\label{sec:discussion}

Here we present our key takeaways while situating our findings in the broader literature. Informed by our observations, we further discuss the tensions and opportunities for individuals, online platforms, and regulatory bodies to improve users' safety against scam-driven human trafficking.

\subsection{Key Insights}

\rev{In the following, we summarize how our key insights extend prior work. In particular, we discover four aspects that influence users' safety and the use of RedNote for exchanging support across multiple phases of scam and human-trafficking.}

\subsubsection{Exploited cultural obligations undermine survivors' safety and well-being from recruitment through reintegration.}
\label{subsec:disc_cultural_obligations}

Although Chinese kinship ties motivate some family members of the trafficking victims to seek help online (\S\ref{sec:post_trafficking_outcomes}), our analysis shows that these same cultural obligations based on kinship ties were exploited at every stage of the trafficking lifecycle---from recruitment to control and reintegration---contrasting prior work that primarily highlights their positive roles in digital bounding and online safety~\cite{lei2025ai, deng-deng25}. During recruitment, traffickers weaponize kinship ties as a recruitment channel and frame refusal as cultural betrayal (\S\ref{sec:target_characteristics}), extending beyond the generic social-tie recruitment strategies documented in prior research~\cite{lazarus2025assessing, luong2025simple}. These obligations thus become intentional vectors of exploitation. The exploitation continues during captivity, where cultural expectations reinforce power imbalances and enable psychological control within scam compounds (\S\ref{sec:exploitation_types}). Reintegration of survivors is further complicated by family rejection (\S\ref{sec:reintegration}), as families often interpret survivors' return as a need for moral correction rather than trauma recovery, akin to the ``victim-offender'' identity dynamics described in criminology research on scammers involved in human trafficking~\cite{Wang2024}.

\add{\subsubsection{Targeted digital skills in recruitment for exploitation.}
We found that digital skills simultaneously increased vulnerability during recruitment and enabled exploitation within scam operations, yet also facilitated survivors' communication during captivity (\S\ref{sec:post_trafficking_outcomes}). This aligns with prior observations~\cite{luong2025simple, jesperson2023trafficking} that traffickers target digitally skilled workers, then repurpose those skills into forced cybercriminal labor~\cite{henderson2023got, hrc2022cyberslaver} (\S\ref{sec:exploitation_types}). Beyond individual-level vulnerabilities highlighted in prior work~\cite{luong2024understanding, lazarus2025assessing, luong2025simple}, our analysis reveals structural vulnerabilities: weak social safety nets, labor precarity, and mobility-driven instability, particularly for left-behind youth and other mobile populations, which create conditions traffickers exploit (\S\ref{subsec:disc_cultural_obligations}). These findings suggest that technical skills alone do not guarantee online or offline safety awareness.}

\add{\subsubsection{Community-based support in light of institutional gaps.}
Prior work on trafficking responses has largely emphasized institutional interventions~\cite{usip2024mekong, ohchr2023-online-scam-sea}, while our findings reveal how RedNote users assemble bottom-up support systems in response to inadequacies in institutional governance. Coordinated interventions succeed sometimes, but they are often slowed by bureaucracy~\cite{gomez2016regime} or undermined when victims are uncertain whether they will face legal punishment or be recognized as victims (\S\ref{sec:post_trafficking_outcomes}). This gap has pushed families toward RedNote for informal information channels and grassroots networks for rescue coordination and ransom verification---channels that offer rapid response times but lack institutional verification mechanisms (\S\ref{sec:institutional_support}).}

\add{Beyond expanding recruitment reach across multiple platforms~\cite{ohchr2023-online-scam-sea, lazarus2025assessing}, RedNote users raised awareness of how traffickers exploit blind spots in cross-border governance by moving victims between multiple Chinese and foreign platforms, making single-platform interventions insufficient (\S\ref{sec:recruitment_approaches}). They also recognized the inefficacy of Chinese communication apps in rescue coordination (\S\ref{sec:institutional_support}). While prior research has examined testimonies of individual survivors~\cite{lazarus2025assessing}, we showed how diverse RedNote users can collectively develop prevention strategies grounded in lived experiences (\S\ref{sec:RQ4}). Survivors shared firsthand accounts and ``confession'' as warnings to the community, and the community generated red-flag lists with hashtags tailored to specific demographics---especially women and younger adults, reflecting RedNote's dominant demographics~\cite{wan-chi25}---and highlighted relevant self-defense strategies (\S\ref{sec:self_defense}).}

\add{\subsubsection{Online visibility as double-edged sword in contested victimhood.}
Platform visibility enables crisis coordination but simultaneously exposes survivors to public judgment about their complicity in fraud. There are cases where users debated whether the survivor knowingly entered the scam work (\S\ref{sec:recruitment_approaches}), perceiving them as ``victim-offenders''~\cite{luong2024understanding, Wang2024}. Survivors must therefore reconcile their own victimization with the harm they caused under coercion, leading many to self-censor preemptively out of fear of judgment. This dynamic contrasts sharply with online community support for unambiguous victims, such as sexual-assault survivors~\cite{pevac2022online}, where disclosure is more likely to evoke empathy than accusations of criminality.}

\add{At the same time, survivors depend on visibility to seek help, posting distress signals and urgent pleas on RedNote. However, the shifting discourse among RedNote users---including slang repurposed from historically discriminatory terms for Chinese immigrant workers~\cite{hui-cswm95}---can re-traumatize survivors (\S\ref{sec:exploitation_types}). Platform visibility also amplifies government warnings by framing voluntary surrender as the only path to leniency (\S\ref{sec:post_trafficking_outcomes}), potentially deterring escape efforts. Our analysis confirms that escape carries severe risks, including abandonment, re-trafficking, or even death~\cite{sarkar2025cyberslavery, luong2024understanding, lazarus2025assessing, ohchr2023trafficking} (\S\ref{sec:post_trafficking_outcomes}), and survivors often face long-term trauma treatment. Fear of judgment and uncertainty about legal consequences thus prevents many from seeking help, compounding the harms they have already endured.}

\subsection{Implications}

Based on our findings, we present the key implications of our work, including tensions that hinder the systematic resolution of scam-driven human trafficking as well as potential venues to mediate these tensions.

\subsubsection{\rev{Structural vulnerability in the labor market}}

The recruitment targets of scam-driven trafficking reflect broader structural vulnerabilities that extend beyond individual risk factors. China's rapid economic development and high penetration rate of Internet services have fostered a massive population of digitally literate people, while widespread digital platform adoption has outpaced safety awareness and protective literacy~\cite{creemers2022china}. Furthermore, geo-political tensions and the COVID-19 pandemic slowed down economic growth and introduced risks in global trade, exacerbating financial pressure and labor surplus on individuals, including skilled workers since 2018~\cite{mckibbin2023global, alessandria2025trade}. Some populations, including left-behind children and unemployed youth identified in our findings (\S\ref{sec:target_characteristics}), still lack access to appropriate support networks and social protections due to urban-rural disparities~\cite{hung2025practice}. Meanwhile, geopolitical shifts in Southeast Asia and the transformation of traditional criminal and drug economies have created new opportunities for traffickers~\cite{jesperson2023trafficking}, creating a vacuum that pulls individuals into recruitment scams. Despite joint efforts between the Chinese government and several affected Southeast Asia countries~\cite{ho2025china}, the threats posed by scam-driven human trafficking are challenging to eradicate due to the complicated economic and geopolitical factors. \add{Rather than relying on technical solutions, which often fall short in the face of such a significant sociotechnical challenge, our recommendations focus on how to better mitigate scams, provide support to survivors, and minimize harms during recovery.}

\subsubsection{\rev{Building culturally and socially informed digital safenets.}}

Our findings reveal the tensions arising from the Chinese social codes and values, which make survivors' families and close contacts a primary coordinator of rescue operations (\S\ref{sec:reintegration}), but it can also be leveraged as a vulnerability in spreading recruitment scams and manipulating survivors (\S\ref{sec:target_characteristics}) or slow down survivor reintegration (\S\ref{sec:reintegration}). These tensions highlight the need to build a safe and trustworthy support network against scam-driven human trafficking, leveraging the Chinese cultural values positively and accommodating the diverse backgrounds of Internet users. One opportunity is to align online safety literacy and education with China's recent push for ``digital technology-empowered grassroots governance''~\cite{liang2022digital_governance}, which mobilizes local resources and personnel to provide timely digital support, including in rural communities. We argue for strengthening ``grassroots governance'' for online safety by engaging trusted community figures in awareness campaigns and family counseling~\cite{wang2007grassroots}. Grassroots initiatives could also adapt proven models such as tech clinics for intimate partner violence survivors in the Western context~\cite{havron-sec19, gupta-sec24} and anonymous legal helplines~\cite{horn2015stop} that help victims and families navigate the legal complications. To support people in remote regions and those working abroad, social media platforms could facilitate the promotion of relevant clinical and counseling services.

    \rev{Furthermore, we hypothesize that individuals vulnerable to scam-driven human trafficking, such as left-behind children, are digitally connected yet remain isolated from positive emotional bonds (\S\ref{sec:target_characteristics}). Such bonds are critical for both preventing recruitment scams and supporting reintegration. Future research could therefore examine the role of culturally informed AI personas as a key component in the safenet, which have shown strong potential for strengthening the digital connections of Chinese users~\cite{lei2025ai, wan-cscw24}. In particular, personified AI agents that highlight social connection and integration beyond individual agency preferred by Chinese users~\cite{sun2025rethinking, ge-chi24} may be capable of providing timely emotional support, safety guidance, and other resources tailored to under-resourced communities. More importantly, we argue that such agents should contribute to mediating kinship relationships within Chinese families rather than merely regulating users' unsafe behaviors. Overemphasis on discipline risks reinforcing power imbalances grounded in filial piety~\cite{deng-deng25}, escalating family tensions, provoking rebellious unsafe online behaviors against Confucian values~\cite{rao2019confucianism}, and ultimately increasing susceptibility to scam and human trafficking risks (\S\ref{subsec:disc_cultural_obligations}).} Nevertheless, deploying AI agents with vulnerable populations requires careful mitigation of potential harm, including privacy risks, over-reliance on AI for critical safety decisions and exploitation by malicious actors \cite{weidinger2022taxonomy}.

\subsubsection{\rev{Improving content recommendation and moderation for safety protection on social media}}

\rev{RedNote hosts a complex ecosystem where warnings, survivor testimonies, grassroots rescue operations, and potentially fraudulent services coexist. Despite considerable efforts to detect and moderate online fraud with law enforcement~\cite{peoplenews25}, our findings show that support information still requires verification to mitigate misinformation and re-traumatization risks. This calls for advances in the application of trauma-informed computing~\cite{chen2022trauma} for content recommendation and moderation systems, by incorporating the cultural and social contexts of scam-driven human-trafficking with foregrounded safety, trust, and survivor agency. First, a key challenge concerns verifying grassroots rescue teams (\S\ref{sec:post_trafficking_outcomes}), which requires collaboration with law enforcement. Second, the success of RedNote's AI-driven scam detection, which intercepted 94.3\% of scam behaviors using over 100 recognition models and reduced user scam reports by 60\%~\cite{peoplenews25}, suggests a pathway to extending these tools for scrutinizing support information moderation.}

\add{Nevertheless, like many other social media platforms, RedNote's recommendation and moderation systems remain opaque~\cite{Ryan2025RedNote}. This opacity and often ``over-sensitive'' moderation lead marginalized users to develop their folk theories~\cite{mayworm2024content, wan-chi25} and adopt coded phrases and repurpose hashtags to evade detection, conceptualized as ``algorithmic resistance''~\cite{Elmimouni2025AlgorithmicResistance} (\S\ref{sec:control_mechanisms}). Moreover, moderation tools risk being weaponized by malicious actors~\cite{sabri2023challenges} and adding pressure on survivors who are already navigating moral debates (\S\ref{sec:institutional_support}). Trafficking-related content poses further moderation challenges due to the contested victimhood. Unlike hate speech, which can often be addressed through standard moderation policies~\cite{zheng-etal-2024-hatemoderate}, harmful discourses questioning survivors' deservingness may go unreported and unmoderated because of self-censorship and moral conflict (\S\ref{sec:public_exposure}). We argue that platforms should shift from passive hosting to active support. They could, for instance, provide tailored, explainable feedback to post authors to improve transparency while maintaining content quality~\cite{jiang2023trade}. Platforms could also deepen collaboration with community experts and leverage existing anti-scam campaigns and volunteer networks~\cite{gaso-website}. These efforts could enhance the precision of moderation, better leveraging human insights with automated moderation tools, and improve the inclusivity and timeliness of online safety content, for example, by highlighting contributions from certified grassroots advocates (\S\ref{sec:post_trafficking_outcomes}) and by offering users dedicated categorization and persistent archives for useful posts (\S\ref{sec:self_defense}).}

\subsubsection{\rev{Overcoming technical barriers in overseas rescue.}}

The current preventive technologies that the Chinese law enforcement deploys focus on domestic network traffic monitoring, blacklisting suspicious applications, and client-side detection and reporting through the \textit{``National Anti-Fraud Center''} app~\cite{verdict2021antifraudapp}. The app was launched in March 2021 by the Ministry of Public Security (China) and is used to detect suspicious calls, messages, and apps, issue warnings, allow users to report fraud, and push prevention content. However, the installation of the ``National Anti-Fraud Center'' app is voluntary~\cite{sohu2023fraud}, and it sees a lack of motivation for potential targets to install it. Additionally, the anti-fraud app, along with other domestic Chinese digital services (e.g., navigation apps), falls short in integrating localized knowledge overseas that is useful in early awareness as well as rescue~\cite{de2017comprehensive}. Law enforcement could consider leveraging trusted local third parties, such as the overseas business associations~\cite{mazerolle2023partnership}, in efforts to build and manage online platforms and communication channels, balancing timeliness of responses and trustworthiness of support.

\subsection{Future Work}

\rev{This study points to several directions for future research. First, while our analysis centers on user narratives on RedNote, future work can examine the \textit{recruitment platforms} themselves. We did not find substantial active evidence that human-trafficking recruitment scams occur through RedNote posts; instead, users primarily use the platform to exchange support. This likely reflects RedNote's strict moderation policies and anti-scam campaigns~\cite{chen2025impact}. However, RedNote users reported encountering scam activities on short-video platforms such as Kuaishou and Douyin (\S\ref{sec:recruitment_approaches}) or via messaging channels (\S\ref{sec:recruitment_approaches}). These observations motivate future investigations into the real-world effectiveness of content moderation and anti-scam interventions across platforms.} 

\rev{Second, future research could evaluate and strengthen \textit{automated detection methods} for scam content to mitigate human trafficking risks, including LLM-based approaches, with particular attention to multi-modal scam contents, evolving scam tactics, and cross-platform scam operations (\S\ref{sec:recruitment_approaches}). Such work must address technical challenges including human--LLM collaboration for scam prevention and reporting~\cite{hossain2025aiintheloop}, the loss of contextual information during platform transitions (\S\ref{sec:recruitment_approaches}), and privacy-preserving detection techniques such as on-device models~\cite{wang2025empowering} or homomorphic encryption~\cite{jin2023fedml}.} 

\rev{Third, future work could track how the \textit{global landscape of scam-driven trafficking} shifts geographically and adapts to anti-scam campaigns and regulatory changes. Cross-platform analyses (e.g., Weibo, Douyin, Facebook, and Telegram) would show how recruitment strategies and resistance narratives unfold under different platform affordances and governance structures. Comparative work across Western, Southeast Asian, and Chinese contexts would further clarify how social, cultural, and technical factors shape the scam and human-trafficking ecosystem, particularly as cross-border scam-driven trafficking expands into Western regions.} 

\rev{Finally, future work could examine how \textit{responsibility is distributed among stakeholders}, including social media platforms, international organizations, and governments, and develop models that support cross-border, multi-stakeholder collaboration. While RedNote's close cooperation with Chinese law enforcement is notable, such frameworks may not generalize to other national contexts. Research should therefore explore alternative support infrastructures (e.g., anti-scam apps, NGO portals, platform reporting tools) tailored to different governance and deployment settings.}

\section{Conclusion}

This study represents the first systematic analysis of scam-driven human trafficking discourse on Chinese social media, examining 158 RedNote posts to understand how users collectively interpret and respond to this emerging form of exploitation. Through qualitative content analysis of recruitment tactics, exploitation mechanisms, reintegration challenges, and protective strategies, we demonstrate how digital communities serve as critical sites for trafficking knowledge construction, survivor testimony, and grassroots resistance efforts. The research reveals important tensions between cultural dynamics that both enable trafficking and complicate survivor support, highlighting opportunities for culturally informed interventions in prevention, platform governance, and cross-border coordination. This research establishes a foundational understanding of how digital communities respond to emerging trafficking forms and identifies opportunities for culturally informed interventions that strengthen prevention efforts, survivor support networks, and cross-border institutional coordination building on existing anti-trafficking frameworks.

\begin{acks}
We thank Tj Elmas and anonymous reviewers for their helpful feedback. We  acknowledge support from the Centre for Doctoral Training in Machine Learning Systems at the University of Edinburgh. The research is partially funded by the Deutsche Forschungsgemeinschaft (DFG, German Research Foundation) under Germany's Excellence Strategy -- EXC 2092 CASA -- 390781972 and by Google through the Google Academic Research Award on Trust and Safety (2024; ID 00029925).
We used a generative AI service, ChatGPT, to assist with light language and style polishing.
\end{acks}

\bibliographystyle{ACM-Reference-Format}
\bibliography{source/main}

\appendix

\section{Keyword Set Used for Crawling}
\label{appendix:keywords}

We list the English translations of key terms used to search and crawl data:

{\ttfamily\small\raggedright
Cambodian scam, Experience of abduction, Experience of trafficked labor, Forced labor, 
Golden Triangle human trafficking, High-paying jobs in Cambodia, High-paying labor, 
Human smuggling, Human trafficking, Human trafficking experience, Human trafficking gang, 
Human trafficking hotspot, Human trafficking survivor, Labor trafficking, 
Myawaddy compound/zone, Myawaddy escape experience, Northern Myanmar, 
Online and offline human trafficking, kidney cutting (organ harvesting), 
Overseas job scam, Recruitment at the Myanmar border, Sharing lessons from overseas, 
Trafficked and trapped overseas, Wage fraud against workers, Working in Southeast Asia, 
Yunnan border defense.
\par}

\section{Excluded Tags}
\label{appendix:excluded_tags}

We list the English translations of the excluded tags:

{\ttfamily\small\raggedright
Classic films and TV, Daily drama-watching, Documentary, Drama chasing, Drama check-in, 
Drama shortage, Food and fun, Good drama recommendations, Good Korean drama recommendations, 
Gourmet / food, Korean drama, Korean drama commentary, Korean drama fan, 
Korean drama recommendation, Korean drama sharing, Local specialties, Movie, 
Movie recommendation, Net worth, New drama updates, No More Bets (film), 
Original work, Popular short drama, Popular TV series, Short drama, 
Show recommendations (when out of ideas), Stress-relief video, TV series, 
TV series recommendation, TV series updates, The drama I'm following, 
Travel recommendation, Trending Korean drama guide, Variety show, 
Watching dramas at home, Web drama.
\par}


\section{Prompt for LLM-based Binary Relevance Filtering}
\label{sec:prompt-llm}

\begin{minipage}{\columnwidth}
\begin{framed}
{\ttfamily\footnotesize
\begin{alltt}
You are Qwen, created by Alibaba Cloud. You are a helpful assistant. 
You will receive a social media post. Please classify the post and 
output 0 for Irrelevant or 1 for Relevant as defined below.

[Topic Definitions]
[0] Irrelevant: The post does not contain specific information about 
scam and human trafficking or cross-border human trafficking (e.g., 
vague and general crime discussions and unrelated lifestyle posts).

[1] Relevant: The post contains specific information about both scam 
and human trafficking or cross-border, particularly in China and 
Southeast Asia, human trafficking (e.g., reports and awareness 
materials about rescue and anti-fraud campaigns, survivors or close 
contacts sharing direct experiences of recruitment scams and 
exploitation, requests for help regarding scams and suspected 
trafficking cases, and prevention strategies, scam awareness tips, 
or warnings against scams or human-trafficking).

Example 1:
\end{alltt}
\begin{CJK}{UTF8}{gbsn}
我前两天去了泰国，然后被人接上了车，在泰国边境的哒府离妙瓦底还有20分钟的车程...到了缅甸妙瓦底器官移植园区...
\end{CJK}
\begin{alltt}
Assignment: 1

Example 2:
\end{alltt}
\begin{CJK}{UTF8}{gbsn}
最近听到网上说泰国旅游很危险，大家最好别去。我朋友说有些地方不安全。
\end{CJK}
\begin{alltt}
Assignment: 0
\end{alltt}
}
\end{framed}
\end{minipage}

\clearpage
\section{Codebook}
\label{sec:codebook}

\begin{table*}[!b]
  \centering
  \caption{Codebook: Recruitment -- Targeting}
  \Description{Four-column codebook table (code, definition, sub-code, sub-code definition) describing recruitment-stage targeting factors. Codes cover lack of awareness (geographical understanding, legal awareness), physical limitation and differences (physical differences, disability/health conditions), over-reliance and overtrust (authority reliance, family/personal relation reliance), financial instability (debt, urgent expenses, uncertain income, ransom), and demographic/social attributes including age, gender, language ability, employment status, marital status, education level, and ethnicity (ethnic minority status; migrant or cross-border ethnic ties).}
  \label{tab:codebook-targeting}
  \includegraphics[width=0.9\textwidth]{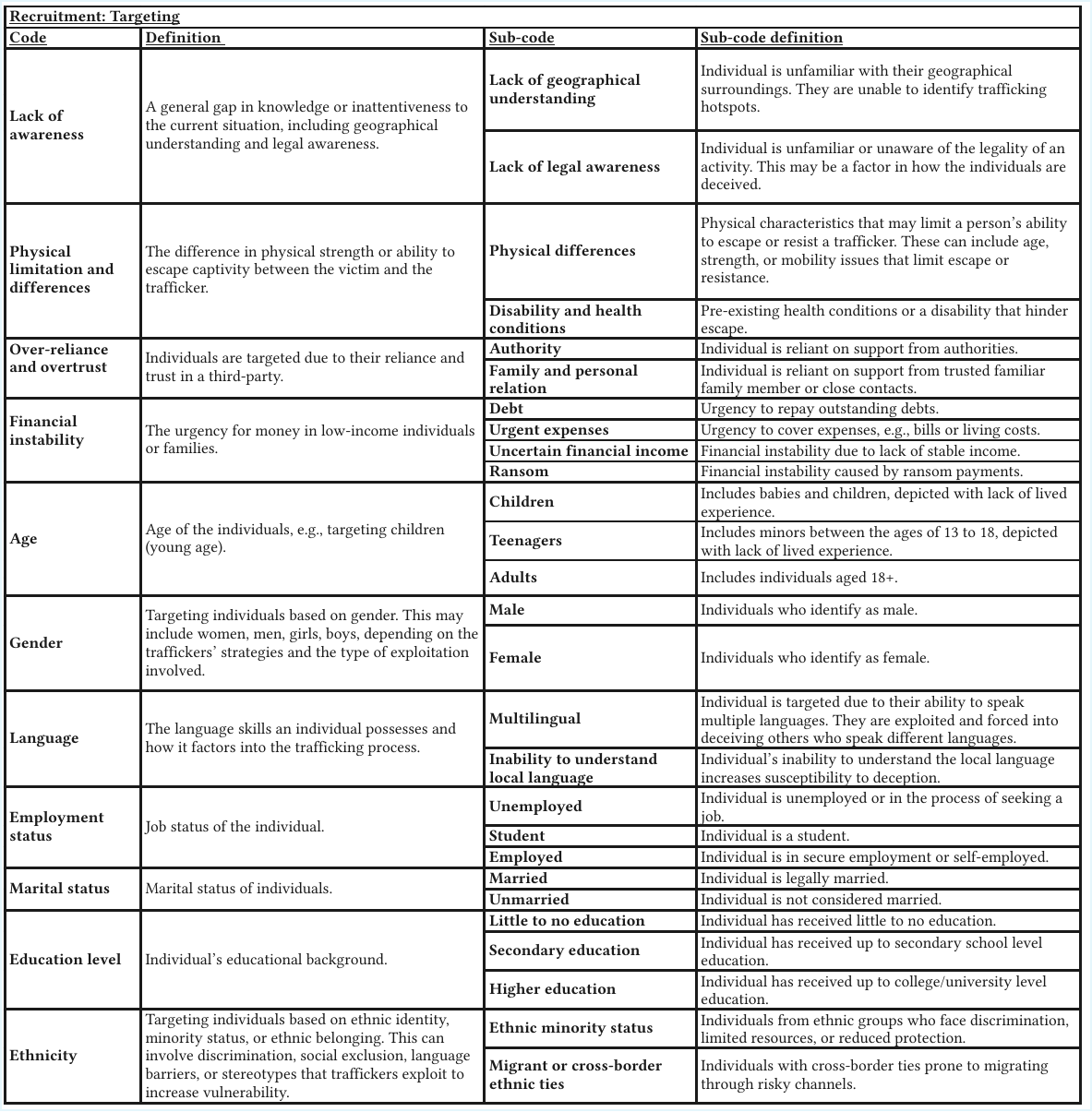}
\end{table*}


\begin{table*}[htbp]
  \centering
  \caption{Codebook: Recruitment -- Recruitment Channels}
  \Description{Four-column codebook table (code, definition, sub-code, sub-code definition) listing recruitment channels used to reach potential victims, including gambling venues, trading and investment platforms, social media platforms, encounters in public spaces, and phone calls. All entries have N/A sub-codes and sub-code definitions.}
  \label{tab:codebook-channels}
  \includegraphics[width=0.9\textwidth]{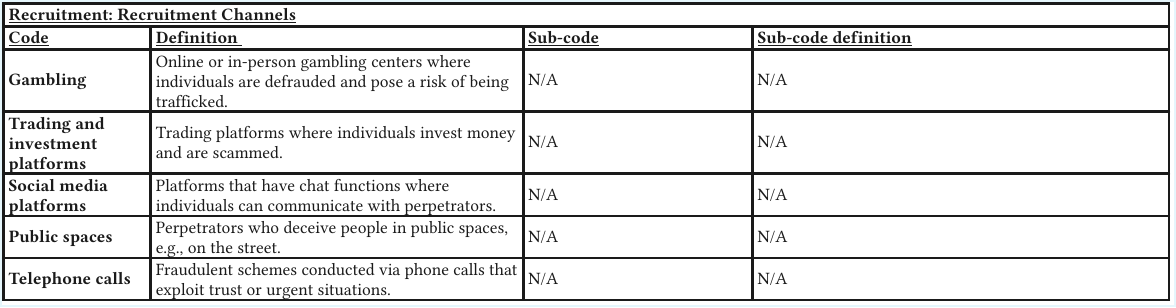}
\end{table*}

\begin{table*}[htbp]
  \centering
  \caption{Codebook: Recruitment -- Recruitment Approaches}
  \Description{Four-column codebook table (code, definition, sub-code, sub-code definition) describing recruitment approaches. Codes include impersonation (authority, service worker, romantic partner, business/service scam, fake family/friends), direct referral (N/A sub-codes), emotional manipulation (distress, family or blood-tie appeals, disability deception), fake promises (foreign/cross-border work opportunity, high-paying job/contracts, travel or studying abroad, high return, upfront rewards such as expense coverage, trial opportunity, advance payment), and travel-based recruitment (voluntary but misled travel; arranged transportation).}
  \label{tab:codebook-approaches}
  \includegraphics[width=0.9\textwidth]{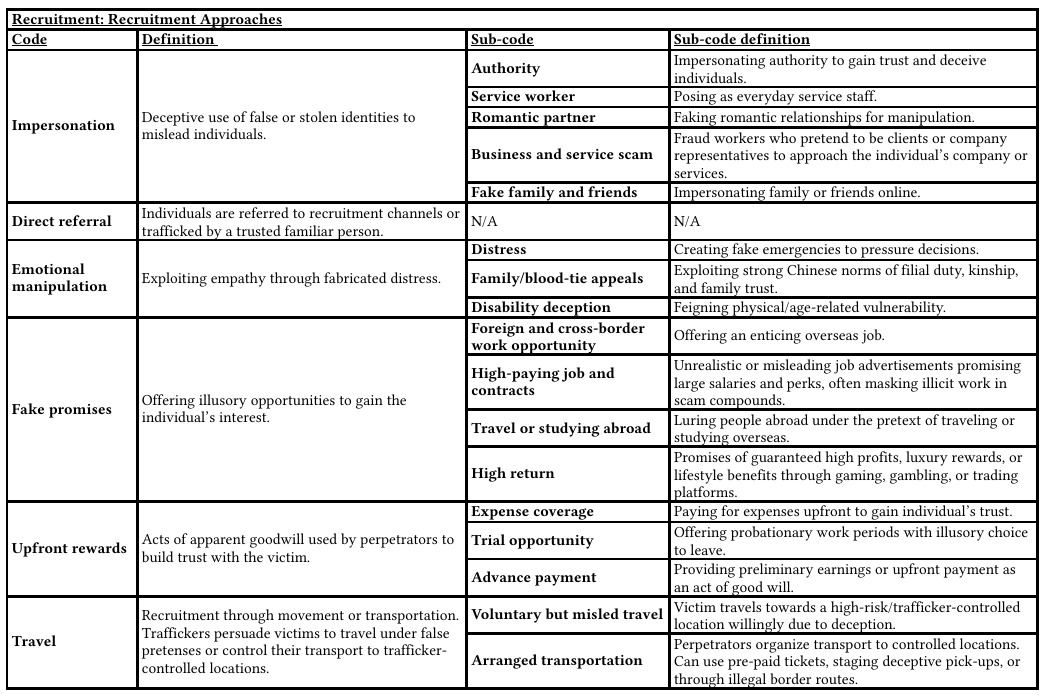}
\end{table*}


\begin{table*}[htbp]
  \centering
  \caption{Codebook: Control Mechanisms}
  \Description{Four-column codebook table (code, definition, sub-code, sub-code definition) defining coercion and control mechanisms during exploitation. Codes include seized documentation and personal possessions; psychological abuse (e.g., threats against family, verbal humiliation, shame and dependency manipulation, brainwashing, fear induction and threat to life); drugs; physical abuse and violence (e.g., beaten, torture, death, rape, electrocution); long work hours; confinement (e.g., locked up, poor compound living conditions, no access to help); and monitoring (e.g., assigned fraud worker, forced searches, surveillance, hunters).}
  \label{tab:codebook-control}
  \includegraphics[width=0.9\textwidth]{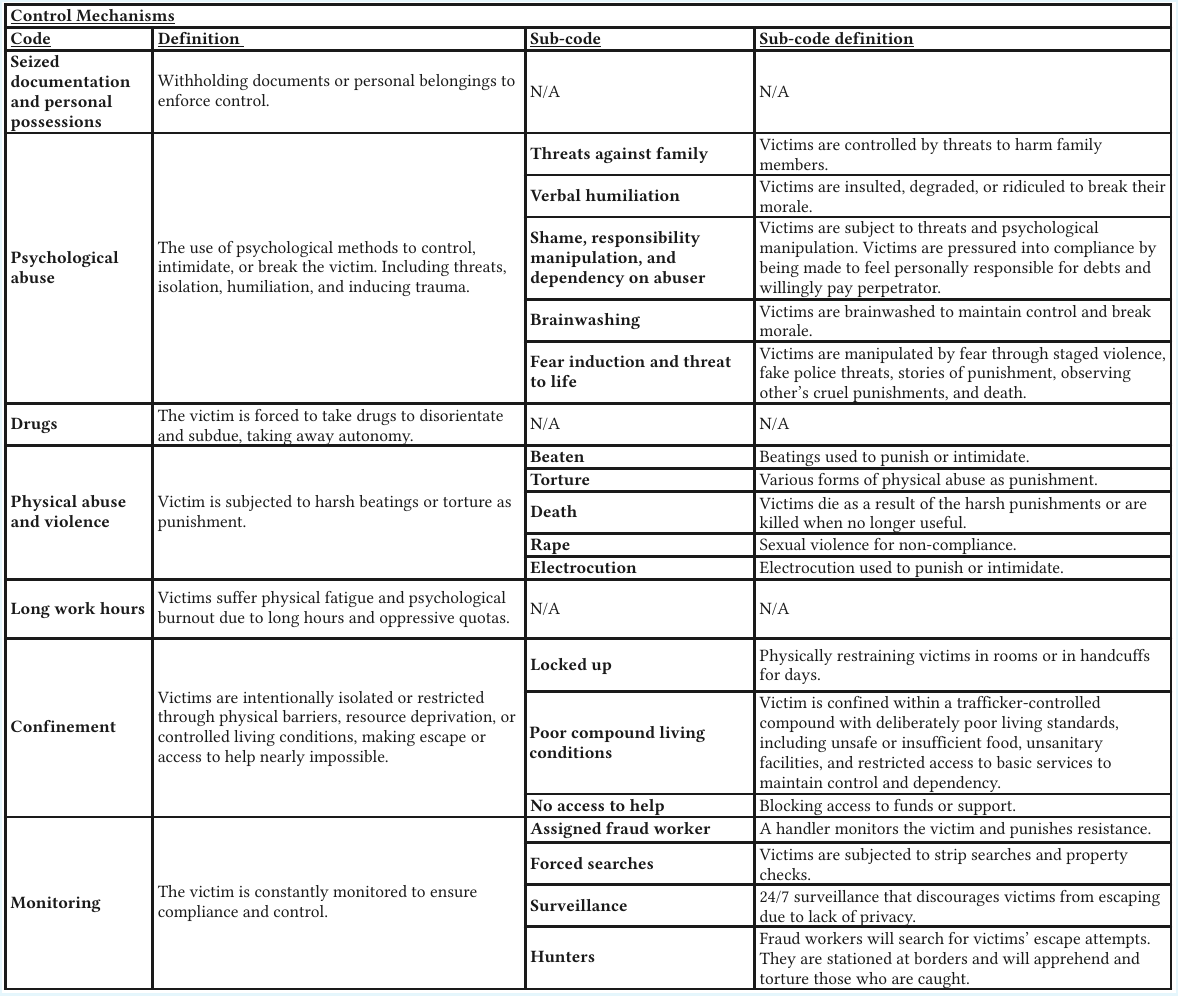}
\end{table*}


\begin{table*}[htbp]
  \centering
  \caption{Codebook: Exploitation}
  \Description{Four-column codebook table (code, definition, sub-code, sub-code definition) defining exploitation type. Codes include fraud-working scheme (telecom fraud and phone scam; romantic fraud; online fraud), sex exploitation (N/A sub-codes), financial exploitation (debt bondage; ransom), organ harvesting (N/A sub-codes), reselling (reselling to scam parks), and child exploitation (N/A sub-codes).}
  \label{tab:codebook-exploitation}
  \includegraphics[width=0.9\textwidth]{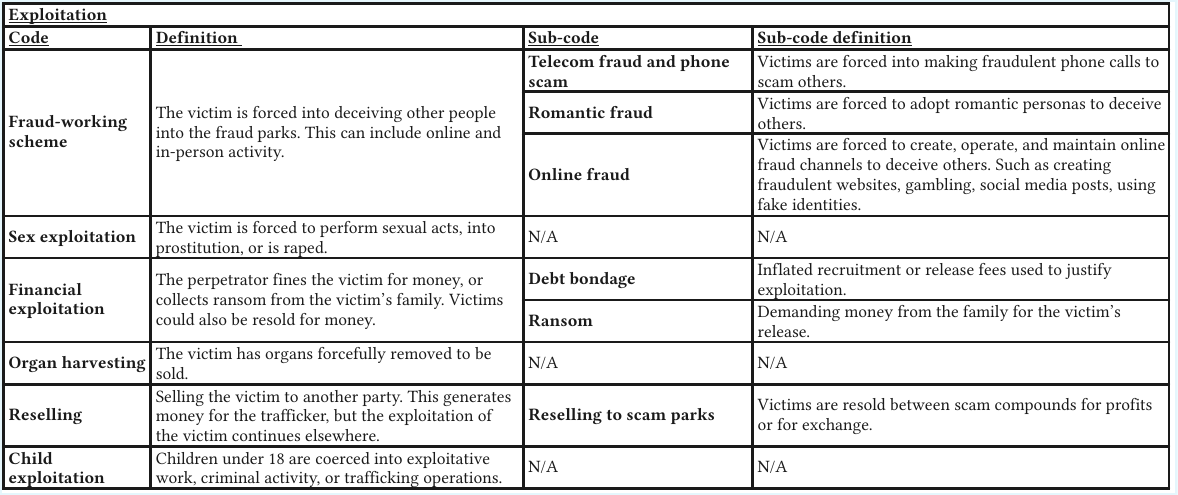}
\end{table*}


\begin{table*}[htbp]
  \centering
  \caption{Codebook: Post-Trafficking -- Outcomes and Risks}
  \Description{Four-column codebook table (code, definition, sub-code, sub-code definition) defining post-trafficking outcomes and risks. Codes include escape (contacting for help; self-escape attempts), rescue (state and international intervention; local authorities intervention; family and friends; grassroots and advocates; failed rescue attempts), abandoned (N/A sub-codes), and retrafficked (scam park transfers; bounty; sold for other exploitation).}
  \label{tab:codebook-post-trafficking}
  \includegraphics[width=0.9\textwidth]{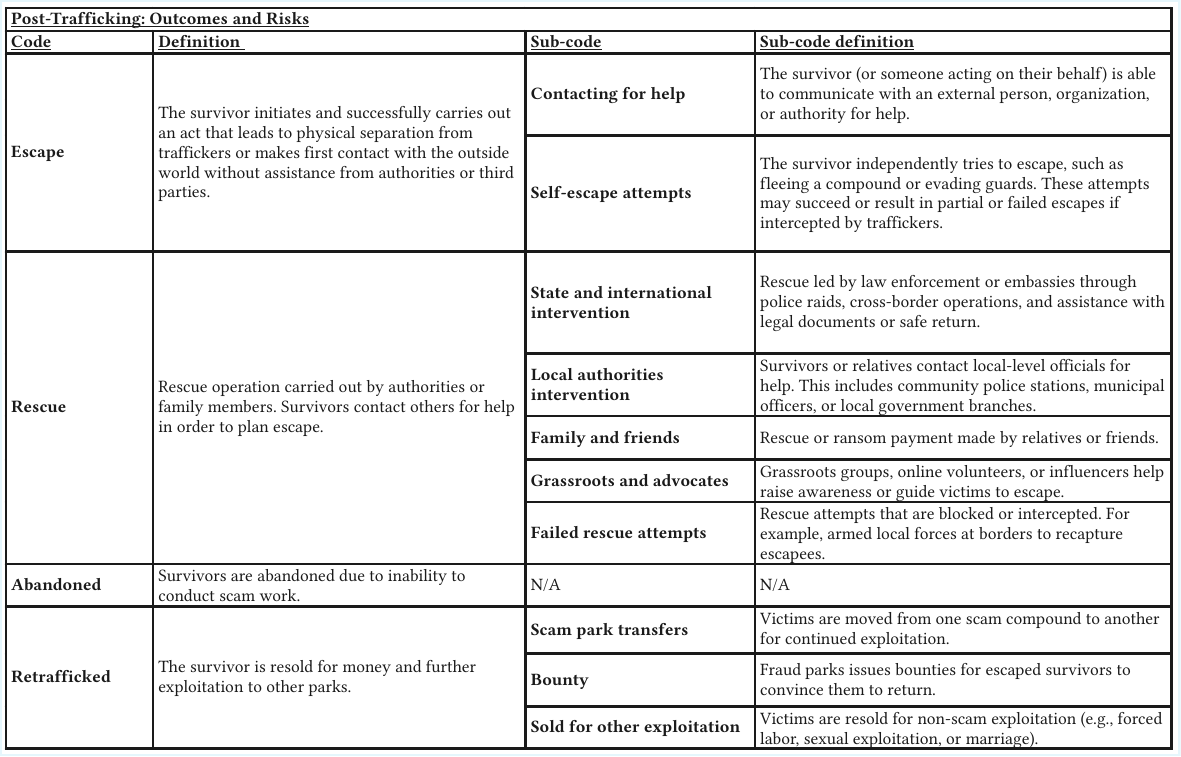}
\end{table*}

\begin{table*}[htbp]
  \centering
  \caption{Codebook: Post-Trafficking -- Support and Reintegration}
  \Description{Four-column codebook table (code, definition, sub-code, sub-code definition) describing post-trafficking support and reintegration. Codes include medical and psychological support (N/A sub-codes), legal procedure (immigration and residency protection; criminal cases; identity and documentation support), and social reintegration (family reunification; experience sharing; community support networks).}
  \label{tab:codebook-recovery}
  \includegraphics[width=0.9\textwidth]{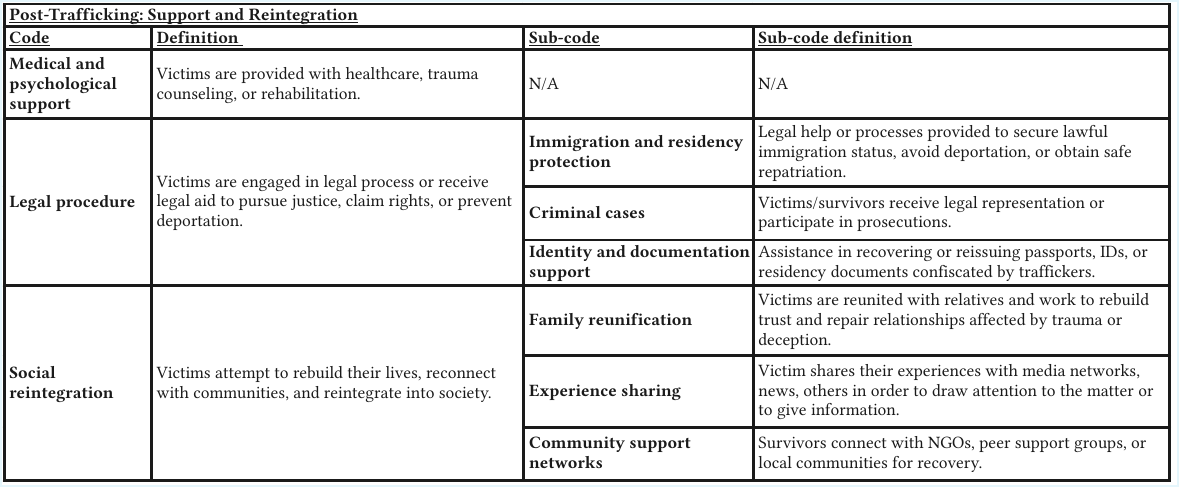}
\end{table*}

\begin{table*}[htbp]
  \centering
  \caption{Codebook: Protective Strategies -- Prevention and Response}
  \Description{Four-column codebook table (code, definition, sub-code, sub-code definition) listing prevention and response strategies discussed online. Codes include female safety tips (N/A sub-codes); legal and labor protection (access legal aid and labor info; avoid illegal travel); caution (N/A sub-codes); indicators of scams (withholding information; third-party apps and messaging platforms); rescue and emergency actions (emergency exit strategies; cooperation; seek help; contact trusted rescue teams); and institutional support (N/A sub-codes).}
  \label{tab:codebook-protective}
  \includegraphics[width=0.9\textwidth]{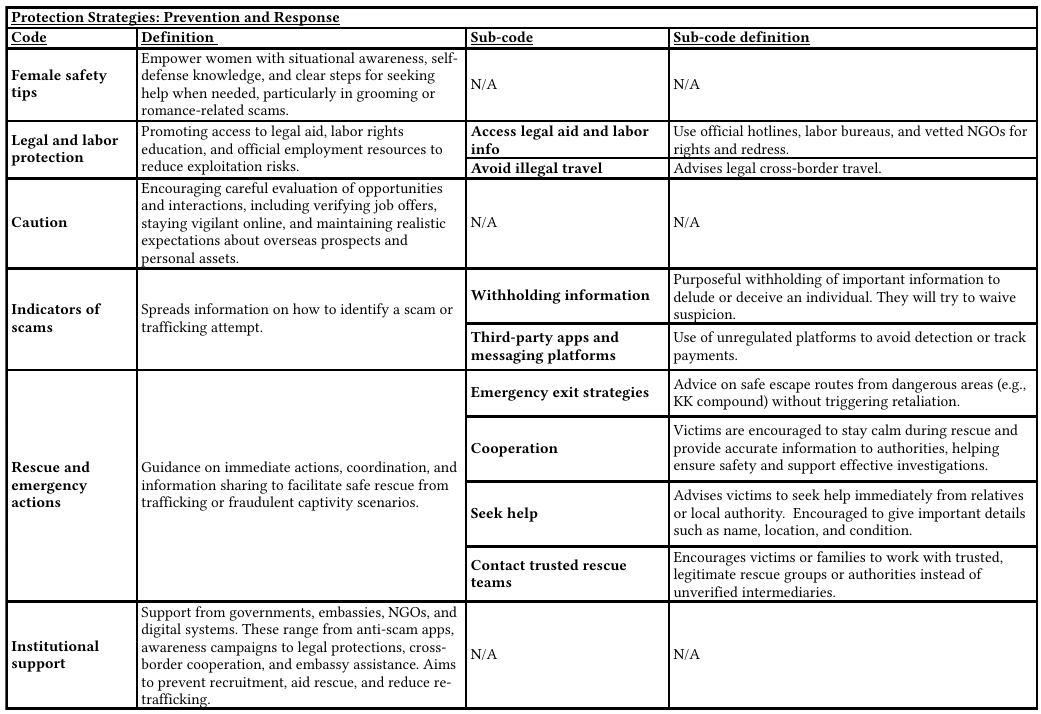}
\end{table*}

\begin{table*}[htbp]
  \centering
  \caption{Codebook: Post Types}
  \Description{Four-column codebook table (code, definition, sub-code, sub-code definition) defining dataset post types: news and awareness, first-person experience, asking for information or support, and tactics against human trafficking. All entries have N/A sub-codes and sub-code definitions.}
  \label{tab:codebook-post-types}
  \includegraphics[width=0.9\textwidth]{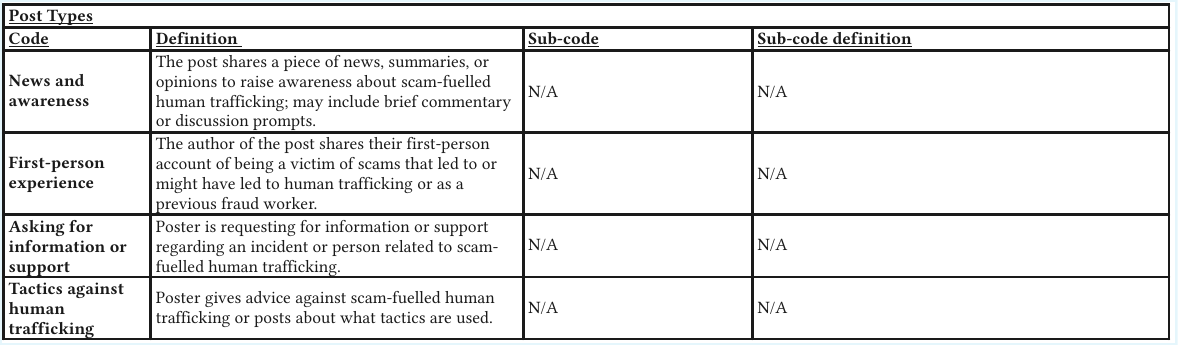}
\end{table*}

\end{CJK}
\end{document}